\documentclass[10pt,noeprint,nourl]{iopart}
\usepackage{graphicx}
\graphicspath{{./},}
\usepackage{bm}
\usepackage{multirow}
\usepackage[table]{xcolor}
\usepackage{transparent}
\usepackage[export]{adjustbox}
\makeatletter
\let\old@makecaption=\@makecaption
\usepackage{subcaption}
\let\@makecaption=\old@makecaption
\makeatother
\usepackage{color}
\usepackage{orcidlink}

\newcommand{\KOsCl}{K$_{\rm 2}$OsCl$_{\rm 6}$}
\newcommand{\KOsBr}{K$_{\rm 2}$OsBr$_{\rm 6}$}
\newcommand{\RbOsBr}{Rb$_{\rm 2}$OsBr$_{\rm 6}$}

\begin{document}

\title{Rotational phase transitions in antifluorite-type osmate and iridate compounds}

\author{A. Bertin$^1$\orcidlink{0000-0001-5789-3178}, L. Kiefer$^1$, P. Becker$^2$\orcidlink{0000-0003-4784-3729}, L.~Bohat\'{y}$^2$\orcidlink{0000-0002-9565-8950}, M.~Braden$^1$\orcidlink{0000-0002-9284-6585}}
\address{$^1$Institute of Physics II, University of Cologne, 50937 Cologne, Germany}
\address{$^2$Sect. Crystallography, Institute of Geology and Mineralogy, University of Cologne, 50674 Cologne, Germany}
\eads{\mailto{bertin@ph2.uni-koeln.de}, \mailto{braden@ph2.uni-koeln.de}}

\begin{abstract}
	
We present temperature-dependent single-crystal diffraction results on seven antifluorite-type $A_2MeX_6$ compounds with $Me$=Os or Ir:  K$_2$OsCl$_6$, $A_2$OsBr$_6$ with $A$=K, Rb, Cs and NH$_4$, and K$_2$Ir$X_6$ with $X$=Cl and Br. 
The structural transitions in this family arise from $MeX_6$ octahedron rotations that generate a rich variety of
symmetries depending on the rotation axis and stacking schemes. In order to search for local distortions in the high-symmetry phase we perform refinements of anharmonic atomic displacement parameters with 
comprehensive data sets.
Even at temperatures close to the onset of structural distortions, these refinements only yield a small improvement indicating only small anharmonic effects.
The phase transitions in these antifluorites are essentially of displacive character. However, some harmonic displacement parameters are very large reflecting
soft phonon modes with the softening covering large parts of the Brillouin zone. The occurrence of the rotational transitions in the antifluorite-type family can be remarkably well analyzed in terms of a tolerance factor of ionic radii.

\end{abstract}
\vspace{2pc}

\submitto{\JPCM}
\maketitle
\ioptwocol

\section{Introduction}

The rich physics of spin-orbit entangled $J=\frac{1}{2}$ states in iridates paves the way to the search of the Kitaev quantum spin liquid~\cite{Kitaev06}, either with Ir ions arranged in a 2D honeycomb lattice~\cite{Chaloupka10} (and related materials~\cite{Banerjee16}) or with 3D honeycomb structures~\cite{Takayama15}. 
Double perovskites of type $A_2BB^{\prime}O_6$ where Ir ions form an $fcc$ lattice became also candidates to host such exotic physics~\cite{Aczel19}. 
The quest continued with iridate antifluorite-type materials, as their $fcc$ lattice can be regarded as a double-perovskite structure with an empty $B^{\prime}$ site. 
In K$_2$IrCl$_6$, which remains cubic down to the lowest temperature and which preserves the ideal octahedral environment at the Ir site, the Kitaev coupling is estimated to be half of the Heisenberg nearest-neighbor exchange coupling~\cite{Khan19}. Despite structural instabilities, the antiferromagnetic ground state of K$_2$IrBr$_6$ is consistent with an extended Heisenberg-Kitaev model with a sizable AFM Kitaev coupling~\cite{Reigiplessis20}. It was argued that the non-cubic crystal-field splitting is small in comparison to Ir oxides preserving the $J=\frac{1}{2}$ ground state~\cite{Khan21}.

The antifluorite compounds of chemical formula $A_2MeX_6$, where the cation $A$ is an alkaline element, $X$ a halide, and $Me$ a four-valent $4d/5d$ transition metal, crystallize in the K$_2$PtCl$_6$ structure type at room temperature, see Figure~\ref{structure_fig}. In contrast to perovskites and related materials,
$MeX_6$ octahedra are not connected by a shared ligand site. Therefore, the coupling of the halide ions is fully asymmetric with a strong 
bond only to a single $Me$ ion. Each halide ion feels thus a steep but asymmetric potential along the bond direction and much flatter ones
in the transversal directions. The antifluorite structure is thus intrinsically unstable against a mostly rigid rotation of the $MeX_6$ octahedra around
an arbitrary axis. 
The absence of shared ligands also results in a weaker coupling between rotations at neighboring sites compared to the
three-dimensional or layered perovskites, where such rotations are tightly coupled. In perovskites, only for a rotation around a bond direction and only
along this direction, the stacking of the rotations can vary. Indeed, numerous rotational phase transitions are reported for members of the  $A_2MeX_6$ family --- see for instance a non-exhaustive list of materials and their associated transition temperatures discussed in terms of cations and halide sizes~\cite{Roessler77}.

The reduced constraints for the rotations in antifluorite-type materials opens the question
about the character of the structural phase transitions. In general structural phase transitions are classified as of order-disorder or of purely displacive
type \cite{Cowley1980}. The rotational transition in SrTiO$_3$ is the textbook example for a displacive transition that is accompanied by the
softening of an associated phonon mode \cite{Cowley1980}, while the rotation of molecular units such as NO$_2$~\cite{Cowley1980} or NH$_4$~\cite{Bruning2020}
can give rise to order-disorder-type transitions. However, there is no clear cut between the two cases and most structural phase transitions
have a mixed character, with a crossover between the two regimes~\cite{Armstrong89,BussmannHolder07}. 
For K$_2$ReCl$_6$ evidence for partial order-disorder character was recently deduced from Raman experiments \cite{Stein2023}.
The atomic displacement parameter (ADP) of the $X$ halide ion perpendicular to the $Me-X$ bond ($U_{\perp}$) is significantly larger than the one parallel to the bond ($U_{\parallel}$), reported for instance in K$_2$IrBr$_6$~\cite{Khan21} or in K$_2$IrCl$_6$~\cite{Khan19}, but so far it remains unclear whether this
enhancement is just a dynamic effect caused by the soft phonon modes, or whether it results from local disorder.
The aim of this work is to quantitatively analyze the role of local distortions in several members of the antifluorite-type family by
refinements of anharmonic models of the ADPs.

As an example, K$_2$ReCl$_6$ undergoes a series of structural transitions understood in terms of octahedra rotations~\cite{Bertin23}. The transitions to a tetragonal and monoclinic space group are characterized by the softening of soft rotary phonon modes at the $X$ and $\Gamma$ points in the Brillouin zone. 
As a consequence of the symmetry lowering, ferroelastic domains emerge and their relation to the magnetic order is at the origin of a huge magnetoelastic anomaly~\cite{Bertin23}. Furthermore, following the work of~\cite{Streltsov20}, this material would constitute an ideal framework to investigate the occurrence of an unconventional Jahn-Teller distortion driven by spin-orbit coupling.

In the case of the $Me$=Os transition metal with $5d^4$ electronic configuration, the ground state is a non-magnetic $J=0$ singlet unless crystal-field effects dominate. The non-magnetic $J=0$ singlet has been confirmed through the measurement of a temperature independent Van-Vleck paramagnetic susceptibility for three different osmate compounds \KOsCl, \KOsBr, and \RbOsBr~\cite{Warzanowski23}. Furthermore, the combination of RIXS and optical-spectroscopy measurements permitted the determination of electronic parameters. Despite undergoing a cubic to tetragonal phase transition, a much weaker non cubic crystal-field splitting, smaller by one order of magnitude compared to K$_2$IrBr$_6$, is reported for \KOsCl, which renders the $J=0$ ground state robust~\cite{Warzanowski23}.

We report here comprehensive single-crystal X-ray diffraction (XRD) studies at temperatures between 250 and 80\,K of five osmate compounds K$_2$OsBr$_6$, K$_2$OsCl$_6$, Rb$_2$OsBr$_6$, Cs$_2$OsBr$_6$, and (NH$_4$)$_2$OsBr$_6$, and of two iridates compounds, K$_2$IrBr$_6$ and K$_2$IrCl$_6$. The emphasis has been laid on the anharmonic treatment of ADPs, in order to characterize the nature of structural phase transitions and the instabilities in this family of materials. In addition we show that the occurrence of the rotational distortion in the broad class of antifluorites can be well understood in terms of a tolerance factor.

\section{Methods}

The antifluorite compounds were synthesized by chemical conversion of H$_2$OsCl$_6$ (respectively H$_2$OsBr$_6$) with the respective alkali metal chloride (respectively bromide) or ammonium bromide and dissolved in diluted HCl (respectively HBr). Small single crystals with a size of ca. 100\,$\mu$m were grown by controlled evaporation of the solvent during a growth period of 2-5 weeks.
The iridate compounds were grown using commercially available K$_2$IrCl$_6$ (respectively K$_2$IrBr$_6$) from diluted HCl (respectively HBr) in a similar way as described above.

Single-crystal XRD measurements have been performed at various temperatures using a Bruker single-crystal diffractometer D8-Venture with  Mo $K_\alpha$ radiation, $\lambda=0.71073$\AA. Reflection intensities have been integrated using the \textit{Apex4} software, and absorption correction and scaling performed with the Multiscale algorithm. Refinements have been carried out with \textit{Jana2020}~\cite{Jana} applying an extinction correction with an isotropic Becker-Coppens formalism~\cite{Becker74}. The quality of each refinements are evaluated from the $R$-factors and GOF parameters provided throughout the main text. Additionally, correlation plots showing the calculated (I(calc)) against the measured (I(obs)) integrated intensities are given in~\ref{correlation_plot}. The sample sizes have been described by a parallelepiped with dimensions reported in Table~\ref{Data_set_all}.

\section{Crystal-structure analyses by X-ray diffraction}
\label{sec_crystal}

\begin{figure}[ht!]
	\includegraphics[width=0.54\columnwidth]{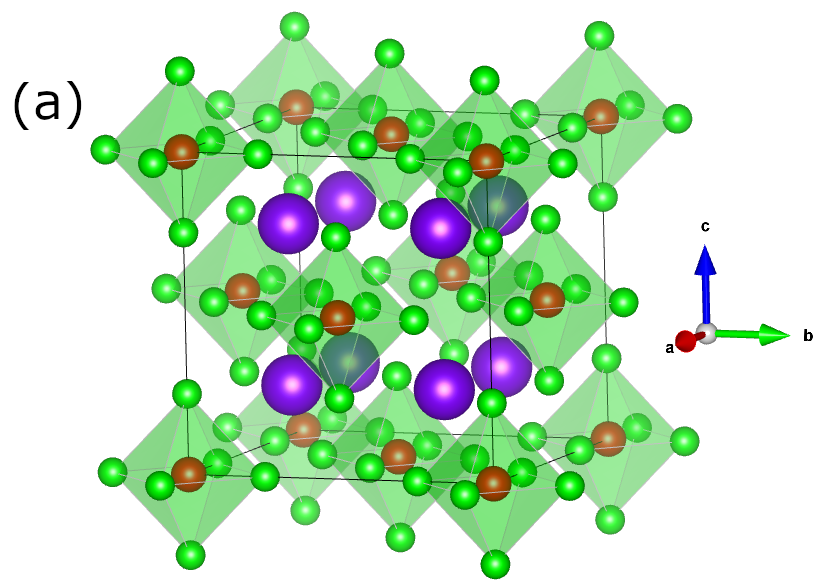}
	\includegraphics[width=0.44\columnwidth]{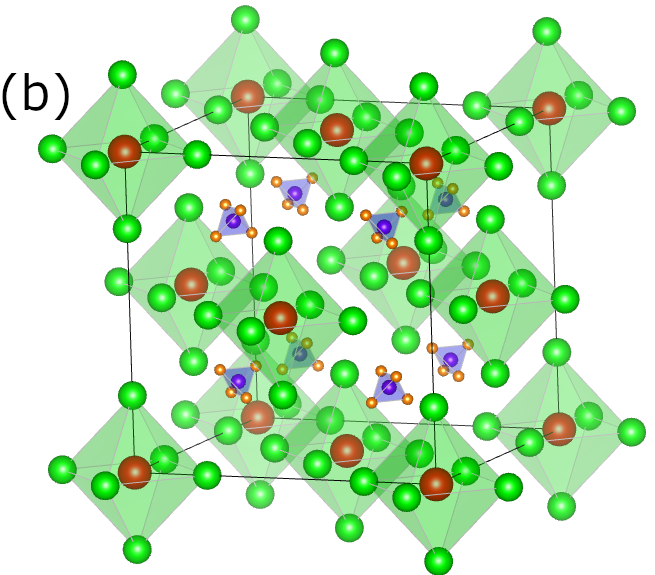}
	\caption{\label{structure_fig} (a) $Fm\bar{3}m$ cubic structure of the antifluorite-type compounds $A_2MeX_6$, where the cation $A$ at  Wyckoff site $8c$ $(\frac{1}{4},\frac{1}{4},\frac{1}{4})$, the transition metal $Me$ at  Wyckoff site $4a$ (0,0,0), and the anion $X$ at Wyckoff site $24e$ ($x$,0,0) are shown with purple, red, and green spheres, respectively. $MeX_6$ octahedra are also highlighted. (b) Same cubic structure for (NH$_4$)$_2$OsBr$_6$ where the tetrahedral ammonium group replaces the cation $A$: N and H ions are shown with purple and orange spheres, respectively. The ammonium group is shown in the {\sl inverted} orientation (model M2, see main text).}
\end{figure}
At room temperature, all samples crystallize in an $fcc$ cubic structure shown in Figure~\ref{structure_fig}(a) and described with the $Fm\bar{3}m$ space group: $Me$ and $X$ ions occupy the Wyckoff positions $4a$ (0,0,0), $8c$ $(\frac{1}{4},\frac{1}{4},\frac{1}{4})$, and $24e$ ($x$,0,0), respectively. In the ammonium complex, N and H ions occupy the Wyckoff positions $8c$ $(\frac{1}{4},\frac{1}{4},\frac{1}{4})$, and $32f$ $(x,x,x)$, respectively. The stoichiometry of the seven samples--- $A_2MeX_6$ with $A=$K, Rb, Cs, NH$_4$ $Me=$Os, Ir and $X=$Cl, Br --- has been verified using the data sets collected at 250K in the cubic phase (and at 200K for K$_2$IrCl$_6$ and (NH$_4$)$_2$OsBr$_6$). 
Refinements with symmetry-allowed harmonic ADPs and with keeping the site occupancy of the transition-metal ion $Me$ to its stoichiometric value have been carried out using the ionic X-ray form factors of K$^+$/Rb$^+$/Cs$^+$, Cl$^-$/Br$^-$, and Os$^{4+}$/Ir$^{4+}$. For the ammonium group, the atomic form factors are used.
The occupancies of the $A$ and $X$ ions have been refined and are tabulated in Table~\ref{stoichiometry}. 
For the $R$-factors and GOF parameters given in Table~\ref{stoichiometry}, values on the left and right sides of symbol ``$/$" correspond to harmonic refinements in stoichiometric and non-stoichiometric conditions, respectively. 
With the exception of Cs$_2$OsBr$_6$, there is no indication for a significant amount of vacancies
or of excess scattering in the Osmate and Iridate samples. While the amount of vacancies at the $A$ site with increasing ionic radius seemingly increases, this minor occupational deviation for Rb$_2$OsBr$_6$ can be ascribed to imperfections of the form factors.
Therefore, we fixed the site occupations to the ideal ones in all refinements described below. With the 250\,K data sets we verified that this does not change any parameters by more than one standard deviation.
Only for Cs$_2$OsBr$_6$ the analysis indicates a small but significant amount of the vacancies at the Cs and Br sites.
 
In the following, crystal structures of the compounds K$_2$OsBr$_6$, K$_2$IrBr$_6$, K$_2$OsCl$_6$, K$_2$IrCl$_6$, Rb$_2$OsBr$_6$, Cs$_2$OsBr$_6$, and (NH$_4$)$_2$OsBr$_6$ will be discussed in details in Section~\ref{KOB_structure}, Section~\ref{KIB_structure}, Section~\ref{KOC_structure}, Section~\ref{KIC_structure}, Section~\ref{ROB_structure}, Section~\ref{COB_structure}, and Section~\ref{NHOB_structure}, respectively. 
\small
\begin{table*}[t!]
	\caption{\label{Data_set_all} Summary of single-crystal data sets completeness and integration parameters for samples Rb$_2$OsBr$_6$, K$_2$OsCl$_6$, K$_2$OsBr$_6$, Cs$_2$OsBr$_6$, K$_2$IrBr$_6$, K$_2$IrCl$_6$, and (NH$_4$)$_2$OsBr$_6$ at all measured temperatures. The corresponding space groups used for the refinements are indicated. Completeness of the data set is denoted through the number of collected/unique/observed reflections.  Unique and observed reflections are averaged within the respective space-group symmetry. The latter group follows the criteria $I/\sigma > 3$. Data have been collected in symmetry P1 with a maximum resolution indicated by $(\sin\theta/\lambda)_{\rm max}$. Note that some data sets have been measured with a smaller resolution $(\sin\theta/\lambda)_{\rm max}$ resulting in a smaller amount of collected reflections, because weak data ($I/\sigma < 2$) become predominant at larger scattering angle $2\theta$. The consistency of every data set is evaluated through the internal weighted $wR^2(int)$ factor after scaling and absorption correction but before averaging reflections with respect to the symmetry of the considered space group. }
	\begin{indented}
		\lineup
\item[]\begin{tabular}{@{}*{9}{lcccccccc}}
			\br 		
			Sample&\multicolumn{2}{c}{Rb$_2$OsBr$_6$}&\multicolumn{2}{c}{K$_2$OsCl$_6$}&\multicolumn{4}{c}{K$_2$OsBr$_6$} \cr
			\mr
			Size&\multicolumn{2}{c}{$52\times78\times78\mu$m}&\multicolumn{2}{c}{$123\times148\times163\mu$m}&\multicolumn{4}{c}{$26\times59\times86\mu$m} \cr
			T (K)&250K&80K&250K&80K&250K&\multicolumn{2}{c}{210K}&80K\cr
			Space group&\multicolumn{2}{c}{$Fm\bar{3}m$}	&\multicolumn{2}{c}{$Fm\bar{3}m$}&$Fm\bar{3}m$&\multicolumn{2}{c}{$P4/mnc$}&$P2_1/n$ \cr
			$N$ reflect.& 39054 & 36537&39652&38209&39293&\multicolumn{2}{c}{139182}&111521\cr
			Unique&181&279&240&239&175&\multicolumn{2}{c}{1665}&6748 \cr
			$(I/\sigma) > 3$&175&268&240&239&172&\multicolumn{2}{c}{988}&2757 \cr
			$(\sin\theta/\lambda)_{\rm max}$&0.83~\AA$^{-1}$&1~\AA$^{-1}$&1~\AA$^{-1}$&1~\AA$^{-1}$&0.83~\AA$^{-1}$&\multicolumn{2}{c}{0.85~\AA$^{-1}$}&0.87~\AA$^{-1}$ \cr
			$a$ (\AA)&10.4265(15)&10.3518(3)&9.7813(12)&9.7144(6)&10.3187(3)&\multicolumn{2}{c}{10.3007(8)$^{\rm a}$}&10.2477(12)$^{\rm a}$\cr 
			$wR^2(int)$&6.12~\%&5.96~\%&6.52~\%&6.75~\%&6.27~\%&\multicolumn{2}{c}{6.75~\%}&8.93~\% \cr	
			
			\mr	
Sample&\multicolumn{2}{c}{Cs$_2$OsBr$_6$}&\multicolumn{2}{c}{K$_2$IrBr$_6$}&\multicolumn{2}{c}{K$_2$IrCl$_6$}&\multicolumn{2}{c}{(NH$_4$)$_2$OsBr$_6$}\cr
\mr
Size&\multicolumn{2}{c}{$18\times22\times30\mu$m}&\multicolumn{2}{c}{$87\times116\times 120~\mu$m}&\multicolumn{2}{c}{$37 \times 44 \times 46 ~\mu$m}& \multicolumn{2}{c}{$32 \times 66 \times 72 ~\mu$m}\cr
T (K)&250K&80K&250K&200K&200K&80K&\multicolumn{1}{c}{200K}&80K\cr
Space group&\multicolumn{2}{c}{$Fm\bar{3}m$}&\multicolumn{2}{c}{$Fm\bar{3}m$}	&\multicolumn{2}{c}{$Fm\bar{3}m$}&\multicolumn{2}{c}{$Fm\bar{3}m$} \cr
$N$ reflect.&31830&21836&48029&130375&92446&95520&\multicolumn{1}{c}{88788}&118931 \cr
Unique&127&127&392&364&336&333&\multicolumn{1}{c}{397}&393\cr
$(I/\sigma)> 3$&122&114 &375&364&336&333&\multicolumn{1}{c}{379}&378 \cr
$(\sin\theta/\lambda)_{\rm max}$&0.71~\AA$^{-1}$&0.71~\AA$^{-1}$&1.14~\AA$^{-1}$&1.11~\AA$^{-1}$&1.14~\AA$^{-1}$&1.14~\AA$^{-1}$&\multicolumn{1}{c}{1.14~\AA$^{-1}$}&1.14~\AA$^{-1}$\cr
$a$ (\AA)&10.6456(7)&10.5761(4)&10.2826(14)&10.2639(3)&9.7245(4)&9.6790(10)&\multicolumn{1}{c}{10.3536(8)} &10.3072(5)\cr
$wR^2(int)$&7.02~\%&7.13~\%&6.09~\%&5.71~\%&6.20~\%&6.76~\%& \multicolumn{1}{c}{5.84~\% }&6.93~\% \cr

\br						
		\end{tabular}
\item[] $^{\rm a}$ Integration with a primitive pseudo-cubic unit cell
		\end{indented}
\end{table*}	
\normalsize

\begin{figure}[ht!]
	\begin{subfigure}{0.49\columnwidth}	
			\includegraphics[width=\textwidth]{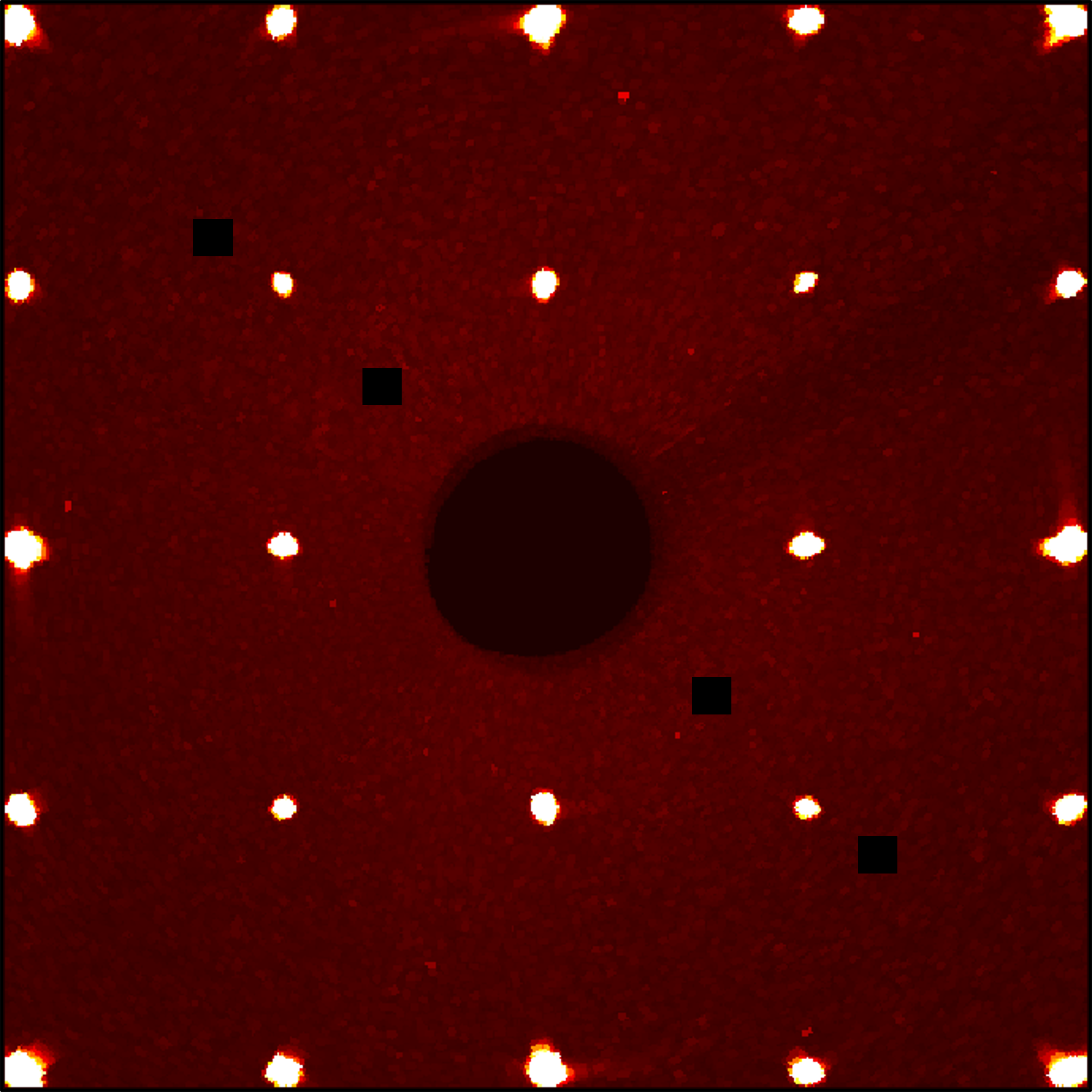}
	\end{subfigure}
	\begin{subfigure}{0.49\columnwidth}
		\includegraphics[width=\textwidth]{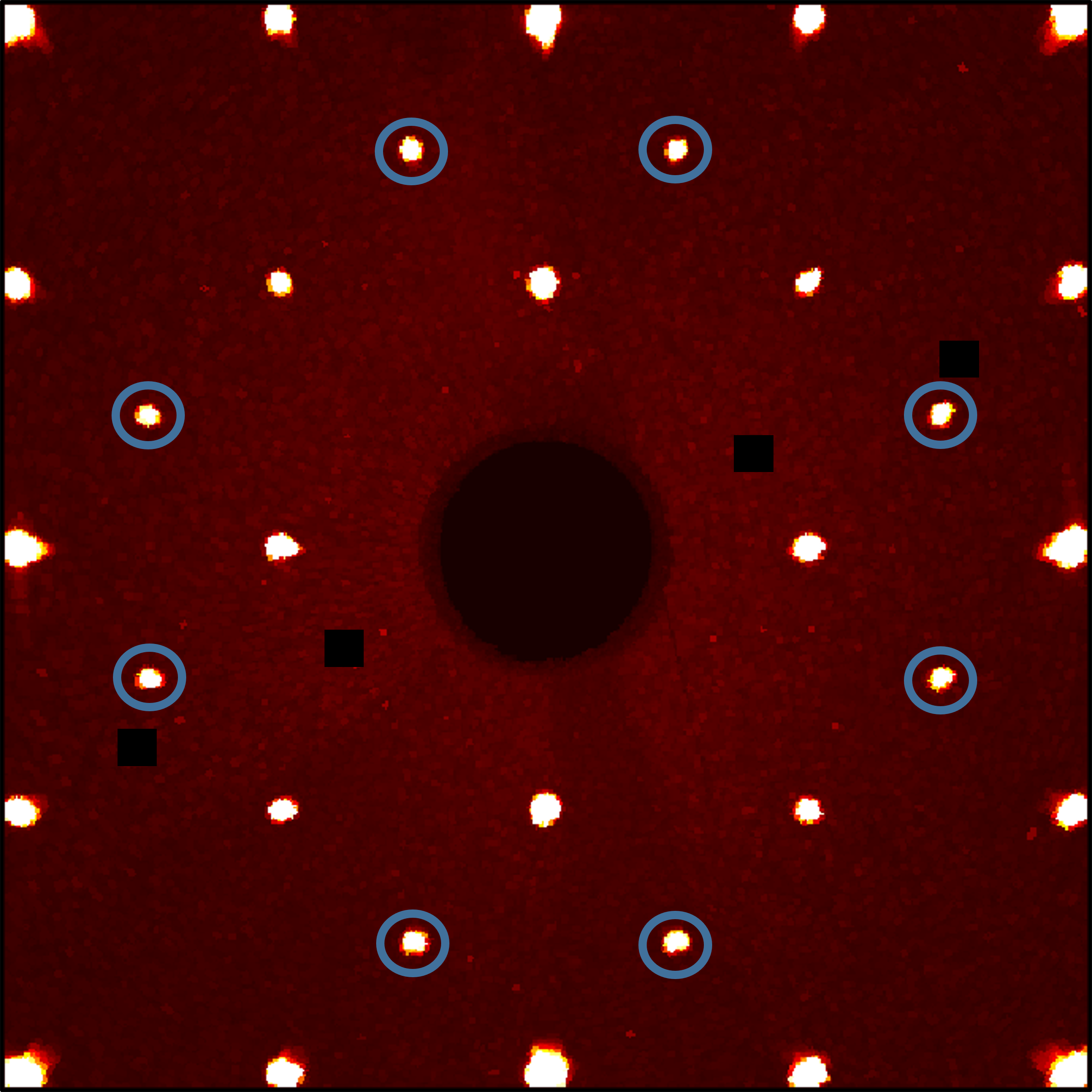}
	\end{subfigure} 
	\begin{subfigure}{0.49\columnwidth} 
		\includegraphics[width=\textwidth]{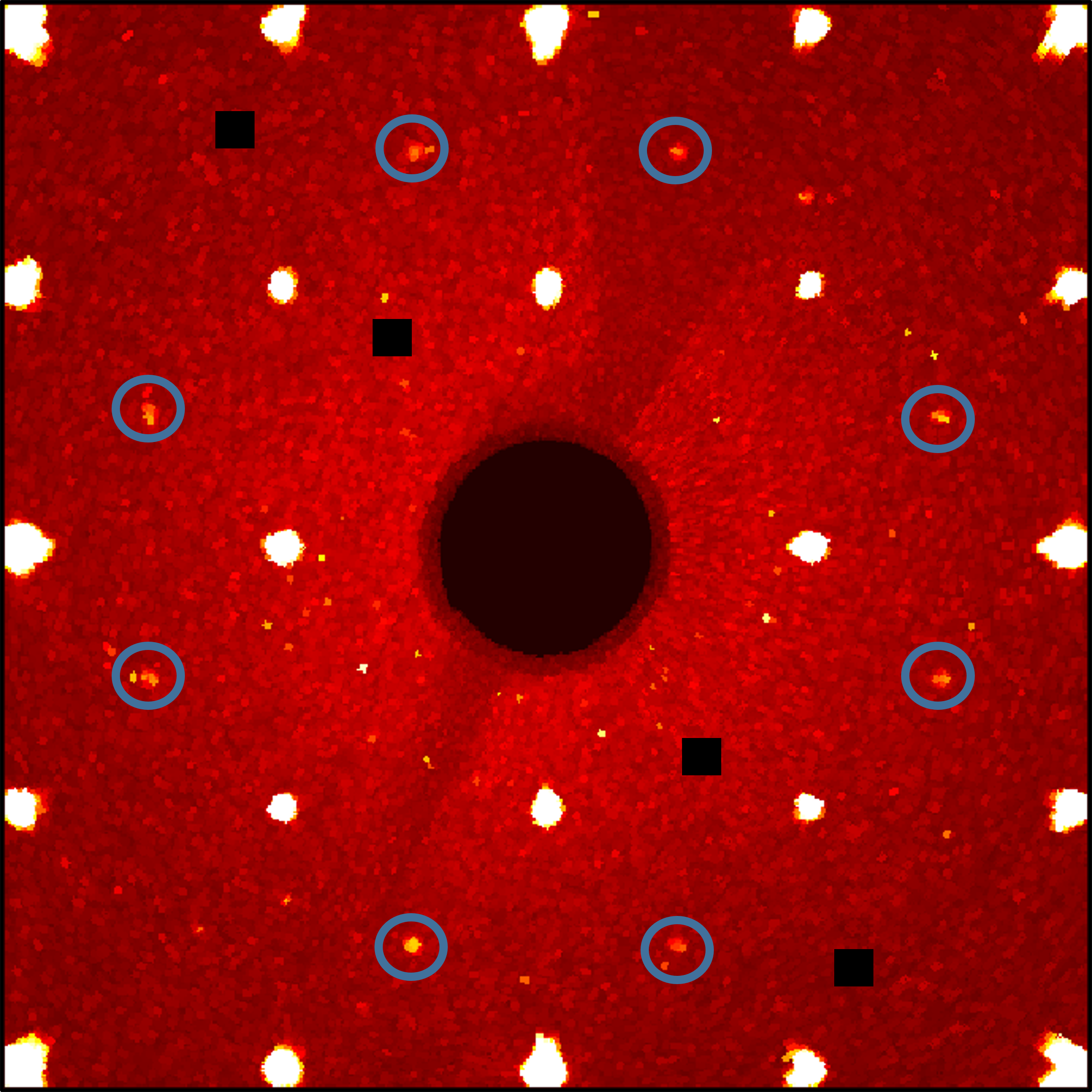} 
	\end{subfigure}  
	\begin{subfigure}{0.49\columnwidth} 
		\includegraphics[width=\textwidth]{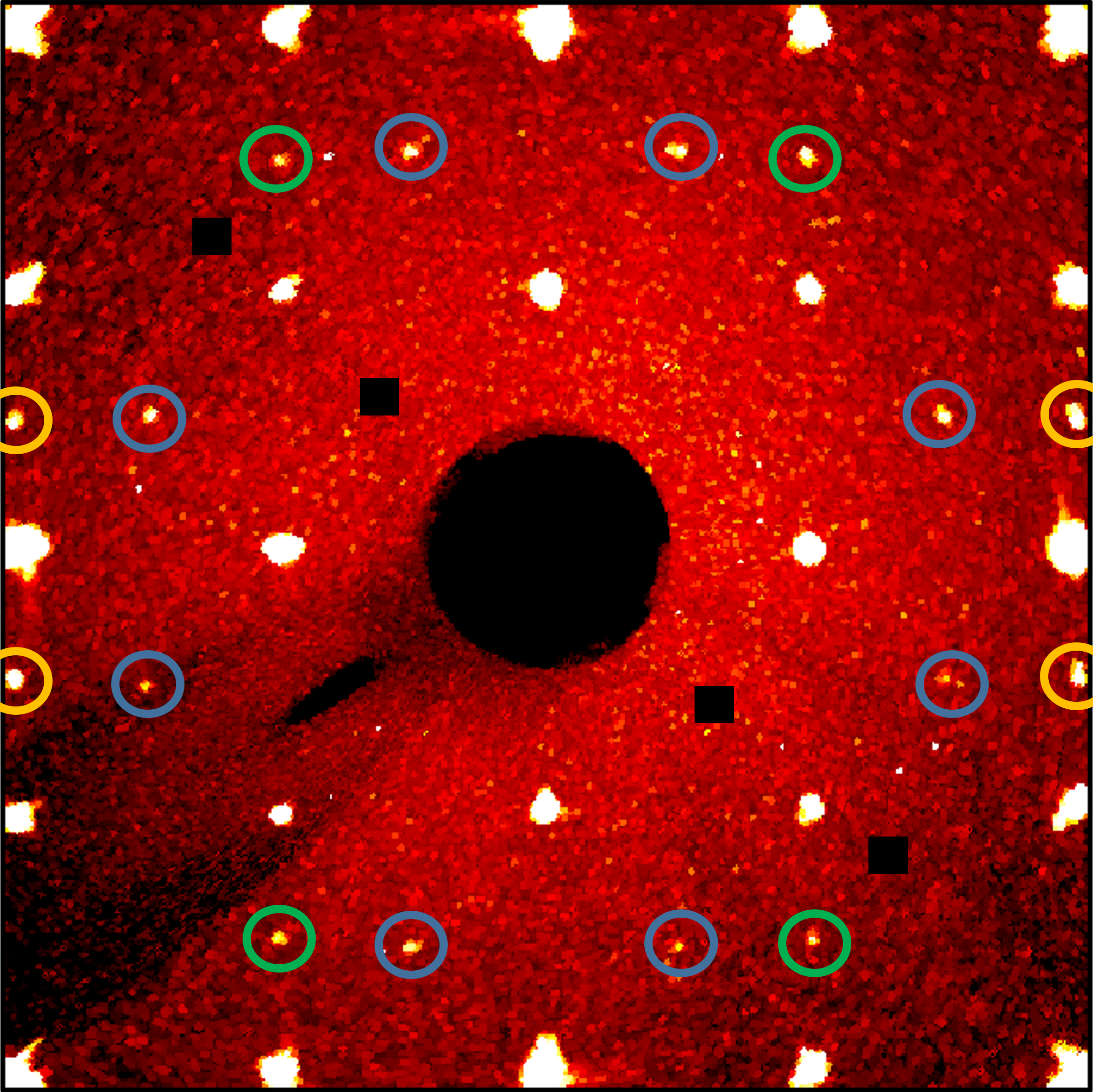} 
	\end{subfigure}
	\makeatletter\long\def\@ifdim#1#2#3{#2}\makeatother
	\caption{\label{KOB_maps}  Calculated precession maps computed for K$_2$OsBr$_6$ with a cubic unit cell. The black squares mask weak spurious reflections from an other grain.  {\sl Top:} The (hk0) reciprocal-space plane at 250K (left) and 210K (right). The blue circles highlight the presence of superstructure reflections of type $(310)$ breaking the $F$ centering. {\sl Bottom:}  (h0l) maps at 210K (left) and 80K (right). The blue circles highlight the presence of superstructure reflections of type $(301)$ breaking the $F$ centering. The green and yellow circles points to superstructure reflections of type $(203)$ and $(401)$, respectively.}
\end{figure}

\small
\begin{table*}[t!]
	\caption{\label{stoichiometry} Summary of the harmonic refinements at 250K in the cubic phase with occupation of the $A$ and $X$ sites being varied (data set measured at 200K for K$_2$IrCl$_6$ and (NH$_4$)$_2$OsBr$_6$). The full occupancy of the Os/Ir ions is fixed in all refinements. Ionic X-ray form factors are used here. The $R$-factors (in \%) and GOF parameters of harmonic refinements in the stoichiometric conditions are reported on the left side of the slash symbol ``$/$", and on the right side refinements with the varied occupancy.}

		\begin{indented}
	\lineup
\item[]
\begin{center}
\begin{tabular}{@{}*{6}{l}}
		\br		
		 $A_2MeX_6$             & Occ.($A$)& Occ.($X$)&$R({\rm all})$&$wR({\rm all})$&GOF(all) \cr             
		  \mr
		K$_2$OsCl$_6$&1.00(2)&1.025(8)&2.17/2.10&2.90/2.85&2.56/2.53 \cr
		K$_2$OsBr$_6$&1.02(1)&1.005(6)&1.34/1.34&1.73/1.72&1.36/1.36 \cr
		Rb$_2$OsBr$_6$&0.981(7)&0.976(6)&1.90/1.77&1.90/1.80&1.56/1.49 \cr	
		Cs$_2$OsBr$_6$&0.953(5)&0.958(6)&1.54/1.14&1.95/1.30&1.48/1.00 \cr		
		(NH$_4$)$_2$OsBr$_6$&0.98(3)$^{a}$&0.998(2)&1.28/1.28&1.43/1.42&1.03/1.03 \cr				
		\mr
		K$_2$IrCl$_6$&1.02(1)&1.015(5)&1.20/1.22&1.83/1.76&1.44/1.39 \cr		
		K$_2$IrBr$_6$&1.01(1)&0.987(4)&2.30/2.35&2.42/2.38&1.53/1.51 \cr				
		\br					
	\end{tabular}
\end{center}	
\item[] $^{\rm a}$ Only the occupancy of the N atom is refined in the ammonium molecule. Atomic form factors are used for N and H atoms.
\end{indented}
\end{table*}
\normalsize

\subsection{K$_2$OsBr$_6$}
\label{KOB_structure}

K$_2$OsBr$_6$ undergoes two structural phase transitions, first at $T_{c,1}=220$K towards a tetragonal symmetry described by the $P4/mnc$ space group, and marked by a pronounced anomaly in specific heat measurements~\cite{SauraMuzquiz22}. The second structural phase transition does not show any pronounced feature in thermodynamic measurements, but synchrotron powder diffraction experiments indicate a further symmetry reduction around $T_{c,2} \approx 200$K to a monoclinic space group $P2_1/n$, where the monoclinic angle $\beta$ starts to deviate from 90$^{\circ}$~\cite{SauraMuzquiz22}.
Therefore, K$_2$OsBr$_6$ exhibits the same sequence of structural phase transitions as the sibling compound K$_2$IrBr$_6$~\cite{Khan21}. Also in the latter compound, fingerprints of the second structural phase transition were not found in thermal expansion or specific heat measurements, but only in powder XRD~\cite{Khan21}.

\begin{figure}[th!]
	\centering
	\includegraphics[width=0.99\columnwidth]{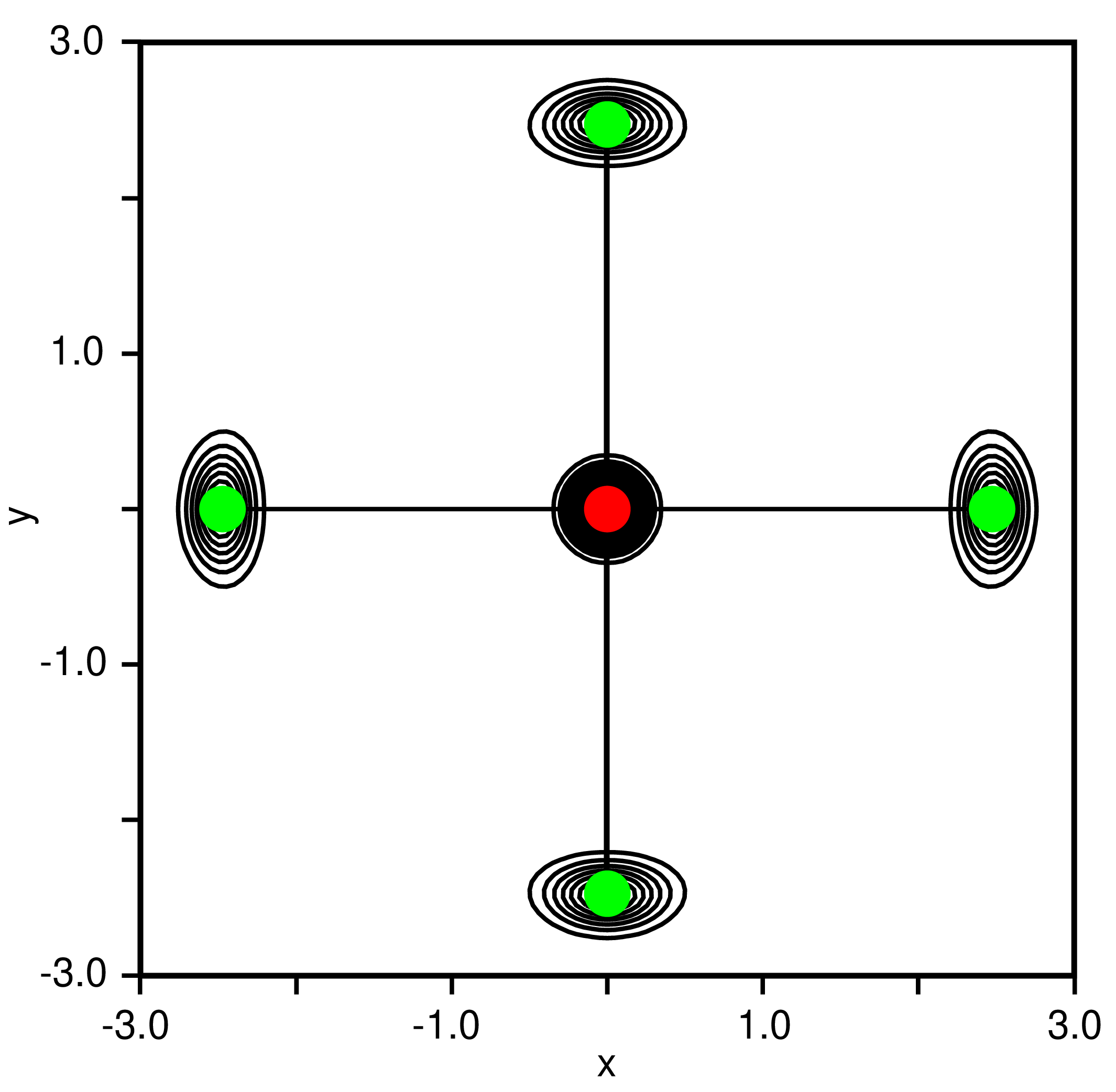}
	\caption{Map of the joint probability distribution function shown in the real space (xy) plane (\AA~units) at $z=0$ obtained in the anharmonic refinement with the data set taken on K$_2$OsBr$_6$ at 250K. The Os and Br ions are shown with red and green spheres, respectively. One recognizes the slight bending of the anisotropic ellipsoids at the Br sites following a rigid rotation.}
	\label{KOB_jpdf}
\end{figure}

Figure~\ref{KOB_maps} shows calculated precession maps computed for K$_2$OsBr$_6$ with a cubic unit cell demonstrating the excellent crystal quality. Few spurious reflections masked by black squares indicate the presence of an additional weakly scattering grain in the sample. The upper two panels of Figure~\ref{KOB_maps} show the (hk0) reciprocal-space planes at 250K in the $fcc$ cubic phase and at 210K in the tetragonal phase. Strong superstructure reflections of type $(310)$  indicated by the blue circles show the breaking of the $F$ centering to a cubic $C$ centered cell with the same lattice parameters. The corresponding space group is $P4/mnc$ or $C4/mcg$ in a non-standard notation preserving the cubic lattice constants. In the tetragonal $C$ centered lattice with $\sim$10\,\AA ~ lattice parameters, the $n$ glide-mirror plane perpendicular to the tetragonal $a_t$ axis with translation component $\frac{b_t}{2}+\frac{c_t}{2}$  becomes a glide-mirror plane perpendicular to the cubic $[110]_c$ axis with translation component $\frac{a_c}{4}+\frac{b_c}{4}+\frac{c_c}{2}$. Because of this unusual translation component, the glide-mirror is labeled $g$ in the non-standard notation used here. The $c$ and $t$ indexes refer to the cubic and tetragonal basis, respectively. In the latter basis, the in-plane $a_t$ and $b_t$ tetragonal axes are shorter by a factor $\sqrt(2)$ and rotated by 45$^{\circ}$ with respect to the in-plane cubic axis. These lattice parameters further correspond to the monoclinic  $a_m$ and $b_m$ parameter in the phase at low temperatures. The lower two panels of Figure~\ref{KOB_maps} show the (h0l) maps at 210K  and at 80K. The blue circles also indicate superstructure reflections of type (301). Since the phase transition is ferroelastic reducing the point-group symmetry from
cubic to tetragonal, three orientations of twin-domains arise with mutually perpendicular orientation of the 4-fold axis.
Therefore, these reflections still preserve the $C$ centering since they arise from $(310)$-type reflections of another twin orientation. 
Note that these superstructure reflections are much weaker pointing to a dominance of one twin orientation, in line with the refinement results at 210K and 80K. Additional superstructure reflections are clearly visible in the (h0l) map at 80K, highlighted by green and yellow circles for $(203)$ and $(401)$-type reflections, unambiguously documenting the breaking of the two $c$ glide-mirror symmetry operations belonging to $C4/mcg$. This is a clear indication of a symmetry lowering phase transition to the monoclinic space group $P2_1/n$, hence supporting the findings of~\cite{SauraMuzquiz22}. This phase transition from tetragonal $4/mmm$ to monoclinic $2/m$ point-group symmetry is of ferroelastic type again and gives rise to further twinning of the crystal. In total, 12 different twin orientations can be expected in the monoclinic crystal at 80 K.

\small
\begin{table*}[th!]
	\centering
	\caption{\label{standard_refinement_KOB} Parameters of refinements with single-crystal XRD data on K$_2$OsBr$_6$ measured in the three different crystallographic phases at 250K, 210K, and 80K. Refinements have been carried out with harmonic and with anharmonic models. The table lines starting with the arrow symbol ``$\rightarrow$" correspond to refinements in the cubic phase with anharmonic terms for the Br ADPs, given in~\AA$^2$,. For the sake of comparison, the free parameters deduced from the anharmonic refinement are listed just below the same parameters deduced from the harmonic refinement. Lattice parameters have been determined by the single-crystal XRD experiments yielding only the average pseudo-cubic parameter and have been taken from Figure\,3 of~\cite{SauraMuzquiz22}. The octahedron rotation angle around the $c$ axis $\phi$ and the octahedron tilt angle $\theta$ around the monoclinic unique axis $b_m$ are calculated. For $\theta$ one may either use the apical Br3 site or the basal plane. Note that the rotation is not strictly around $b_m$.   Since in~\cite{SauraMuzquiz22}, data at 80K are not available, we use the lattice parameters at the lowest available temperature, i.e.\ 90K. The error bars have been estimated from the symbol sizes. The octahedra remain undistorted across the different structural phase transitions.  $twvol$ gives the volume portion of the ferroelastic twins.
	}
		\begin{indented}
	\lineup
	\item[]
	\begin{center}
	\begin{tabular}{@{}*{4}{l}}
		\br  
		&250K&210K&80K \cr
		&$Fm\bar{3}m$&$C4/mcg$ &$P2_1/n$	\cr
		&&($P4/mnc$)& \cr
		\mr
			This work$^{\rm a}$&a=10.3187(3)\AA&a=10.3021(8)\AA&a=10.2477(12)\AA \cr
		\mr
		\multirow{4}{*}{Reference \cite{SauraMuzquiz22}}&\multirow{4}{*}{$a=10.307(2)$\AA} &\multirow{2}{*}{$a=7.244(1)$\AA}&$a=7.186(1)$\AA\cr
		&              &  &$b=7.189(1)$\AA \cr
		&              &\multirow{2}{*}{$c=10.382(2)$\AA}                &$c=10.402(2)$\AA \cr 
		&          &  &$\beta=90.26(1)^{\circ}$\cr   
		\mr
		Os $(x,~y,~z)$&(0,0,0)&(0,0,0)&(0,0,0)\cr
		$U_{\rm iso}$(Os)&0.01835(9)&0.0180(1)&0.0131(1) \cr			
		$\rightarrow$&0.01837(7)&& \cr
		K $(x,~y,~z)$&($\frac{1}{4},\frac{1}{4},\frac{1}{4}$)&($\frac{1}{4},\frac{1}{4},\frac{1}{4}$)&0.4912(3), 0.0303(3), 0.2481(6)\cr
		$U_{\rm iso}$(K)&0.0461(4)&0.0423(5)&0.0177(4) \cr	
		$\rightarrow$&0.0459(3)&& \cr		
		Br1 $(x,~y,~z)$ &0.24028(4), 0, 0&0.24087(6), 0.02294(9), 0&0.2740(2), -0.2092(3), 0.0192(1)\cr
		$\rightarrow$	&0.23993(4)&& \cr
		Br2 $(x,~y,~z)$  &0, $x$(Br1), 0&$-y$(Br1), $x$(Br1), 0&0.2102(3), 0.2732(2), 0.0128(2)\cr			
		Br3 $(x,~y,~z)$  &0, 0, $x$(Br1)&0, 0, 0.2388(1)&-0.0349(1), 0.0030(2), 0.2380(1)\cr				
		$U_{\parallel}$ $U_{\perp}$ (Br)&0.0182(2)~0.0590(2)&0.0185(2)~0.0444(2)&0.0141(1)~0.0167(1) \cr	
		$\rightarrow$ &0.0190(3)~0.0585(2)&& \cr									  								
		\mr	
		twvol 1,2,3	(\%)	&-&93.4(2),~2.8(1),~3.8(1)&18.3(7),~16.4(2),~2.4(2) \cr	
		twvol 4,5,6	(\%)	&-&-&1.0(2),~1.1(2),~0(f) \cr	
		twvol 7,8,9	(\%)	&-&-&0.4(2),~1.1(2),~29.8(3) \cr	
		twvol 10,11,12 (\%)		&-&-&0.7(2),~27.5(2),~1.5(2) \cr	
		\mr
		$\phi$&-&$5.4(1)^{\circ}$&$7.5(1)^{\circ}$	\cr
		$\theta_{apical}$,$\theta_{basal}$&-&-&$6.0(1)^{\circ}$, $5.4(1)^{\circ}$ \cr
		\mr
		$R$, $wR$ (obs)(\%)&	1.29~1.61&4.96~5.80&6.17~6.36	\cr
		$\rightarrow$       &1.02~1.14&& \cr
		$R$, $wR$ (all)(\%)&1.34~1.64&7.38~5.95&15.43~6.84	\cr
		$\rightarrow$       &1.07~1.18&& \cr	
		GOF (obs/all)&1.27~1.28&3.56~2.81&2.67~1.83	\cr	
		$\rightarrow$       &0.91~0.93&& \cr
		\br				
	\end{tabular}
	\end{center}
\item[] $^{\rm a}$ For the lower symmetry phases, a primitive pseudo-cubic cell is used for the data integration.
\end{indented}	
\end{table*}	
\normalsize

\small
\begin{table*}[t!]
	\centering
	\caption{\label{anharmonic_param_all} Summary of the symmetry-allowed anharmonic ADPs  of $X$ ions refined at 250K and 80K, for Rb$_2$OsBr$_6$, K$_2$OsCl$_6$, K$_2$OsBr$_6$, and Cs$_2$OsBr$_6$. The dimensionless coefficients $C_{ijk}$ and $D_{ijkl}$ are multiplied by $10^3$ and $10^4$, respectively. Some parameters are set to 0 [resulting in a vanishing error indicated by (f)] when the refined standard deviation exceeds the refined absolute parameter value.}
		\begin{indented}
	\lineup
	\item[]
	\begin{center}
	\begin{tabular}{@{}*{8}{l}}
		\br	
		&&$C_{111}$&$C_{122}=C_{133}$&$D_{1111}$&$D_{1122}=D_{1133}$&$D_{2222}=D_{3333}$&$D_{2233}$\cr
		\mr	
\multirow{1}{*}{K$_2$OsBr$_6$}		&250K&0(f)&-0.00048(5)&0(f)&0(f)&0(f)&-0.00021(5) \cr	
\mr			
\multirow{2}{*}{K$_2$IrBr$_6$}		&250K&0(f)&-0.00043(4)&0.00012(3)&0.00004(1)&0.00035(7)&-0.00016(4) \cr
&200K&0(f)&-0.00031(4)&0.00009(3)&0.00005(1)&0.00008(7)&-0.00021(4) \cr			
		\mr
\multirow{2}{*}{K$_2$OsCl$_6$}		&250K&0(f)&-0.00002(1)&0(f)&0(f)&0(f)&-0.00015(9) \cr
&80K&0.0002(1)&0(f)&-0.00037(9)&-0.00009(2)&-0.00028(8)&-0.00007(5) \cr
\mr	
\multirow{2}{*}{K$_2$IrCl$_6$}		&200K&0(f)&-0.00026(5)&0(f)&0(f)&0(f)&0(f) \cr
& 80K&0(f)&0(f)&-0.00011(4)&-0.00005(1)&-0.00013(4)&-0.00004(2) \cr
		\mr
		\multirow{2}{*}{Rb$_2$OsBr$_6$}	&		250K&0(f)&-0.00017(4)&0(f)&0.00008(2)&0.00042(7)&0.00007(3) \cr
		&	80K&0(f)&-0.00007(3)&0.00006(4)&0.00002(1)&0.00018(4)&0.00003(2) \cr
		\mr			
		\multirow{2}{*}{Cs$_2$OsBr$_6$}		&250K&0.0006(2)&0.00013(8)&0.0003(2)&0(f)&0(f)&-0.00007(6) \cr
&80K&0.0003(2)&0.00016(9)&0.0003(2)&0.00009(5)&0.0002(1)&0(f) \cr
\mr
		\multirow{2}{*}{(NH$_4$)$_2$OsBr$_6$}	&250K&0(f)&-0.00010(1)&0(f)&0(f)&0.00008(2)&0(f) \cr
												&80K&0(f)&0(f)&0(f)&0(f)&0.000024(9)&0(f) \cr
\br
	\end{tabular}
\end{center}	
\end{indented}	
\end{table*}
\normalsize

\small
\begin{table}[h!]
	\centering
	\caption{\label{KOB_250K_order_disorder} Refinements of the structural model with split Br site with the single-crystal XRD data measured at 250K on K$_2$OsBr$_6$  (space group $Fm\bar{3}m$).}
		\begin{indented}
	\lineup
	\item[]\begin{tabular}{@{}*{5}{l}}
		\br 
		&Occ. ($24e$)&x&y&z \cr
		Br&1&0.24027(4)&0.0158(9)&0\cr			
		\mr
		ADPs   & $U_{\rm iso}$ (Os) &$U_{\rm iso}$ (K) &$ U_{\parallel}$ (Br) &$ U_{\perp}$ (Br)   \cr	
		 (\AA$^2$)&0.01830(9)&0.0461(4)&0.0184(2)&0.0446(17) \cr
		\mr		
		$R$-values & $R({\rm obs})$&$wR({\rm obs})$&$R({\rm all})$&$wR({\rm all})$\cr
		 (\%)& 1.24&1.54&1.29&1.57 \cr
		&& GOF(obs)&GOF(all)& \cr    
		&&1.22&1.24& \cr	 
		\br  			
	\end{tabular}
\end{indented}	
\end{table}
\normalsize

Single-crystal XRD data have been collected in the three different crystallographic phases: at 250K in the $Fm\bar{3}m$ phase, at 210K in the tetragonal $P4/mnc$ phase, and at 80K in the $P2_1/n$ phase. An overview of the data completeness is given in Table~\ref{Data_set_all}. 
Because of the small tetragonal and monoclinic distortion, and because of the twinning inherent to such symmetry lowering, both data sets have been integrated with a pseudo-cubic primitive unit cell with lattice parameters given in Table~\ref{Data_set_all}. 
The refined parameters in each phase are summarized in Table~\ref{standard_refinement_KOB}. 
For the cubic phase, the excellent $R$-values and GOF parameters indicate the high sample quality. 
Within the cubic symmetry, ADPs of Os and K ions stay isotropic, while two anisotropic ADPs are introduced for the Br ions: $U_{\parallel}$ and $U_{\perp}$ describing the atomic displacement parallel and perpendicular to the Os-Br bond, respectively. 
For the refinements in the lower symmetry phases, ADPs have been constrained to the symmetry allowed ADPs of the cubic phase. Therefore, ADPs of Os and K ions are kept isotropic, and only two harmonic ADPs for the Br ions are used at all temperatures. As it is commonly observed in the antifluorite-type compounds with disconnected $MeX_6$ octahedra, the transverse displacement of Br ions with regard to the Os-Br bond is about three times larger than that parallel to the bond. Soft rotational modes yield strongly anisotropic ADPs due to their large thermal occupation resulting from the low energies, as it has been quantitatively analyzed in La$_{2-x}$Sr$_x$CuO$_4$~\cite{Braden01}. 
In the antifluorite-type structure, which can be considered as a double perovskite with one empty site, the rotating octahedra are not connected by shared ligands, which
yields a more local character of the rotations and thus a flatter dispersion. The impact of the phonon softening on the ADPs is much enhanced, because the softening occurs in a large fraction of the Brillouin zone.

As expected, all ADPs decrease monotonously with decreasing temperature, and only in the $P2_1/n$ phase, the Br ADPs become almost isotropic. In the tetragonal phase the rotation around only one direction becomes static, while in $P2_1/n$ there are static rotations around three orthogonal directions.
In the tetragonal phase, the three domain volumes have been refined, and indicate an almost untwinned sample with a single dominating domain ($\approx 93$\%). In the monoclinic phase, with now twelve domains to consider, we also find that only 1/3 of the possible domains need to be considered in the refinement, in line with the refinement results in the tetragonal phase.

Even though the structural transition from cubic to tetragonal in K$_2$IrBr$_6$ is driven by octahedra rotations, the hysteretic behavior of thermal expansion measurements and the phase coexistence revealed by powder XRD indicate a first-order character, possibly caused by a tetragonal strain~\cite{Khan21}.
In this case, we may expect very strong anharmonicity of the Br ADPs and even a split probability distribution function. Therefore, a refinement of anharmonic ADPs with the data set collected at 250K in the cubic phase has been conducted. The refined parameters are summarized in Table~\ref{standard_refinement_KOB}. 
The table lines starting with the symbol ``$\rightarrow$" indicate the values obtained in the anharmonic model.
The refined anharmonic ADPs are tabulated in Table~\ref{anharmonic_param_all}. The $R$-values and GOF parameters are significantly improved yielding two anharmonic contributions exceeding at least four times their errors: $C_{122}=C_{133}=-0.00048(5) \times 10^{-3}$ and $D_{2233}=-0.00021(5) \times 10^{-4}$. 
However, we must emphasize that the refinement quality is already very high without the anharmonic terms, which just improve a little bit the description of the detailed probability distribution, which is shown in Figure~\ref{KOB_jpdf}. In this map one sees the bending of the probability ellipsoid at the Br sites that is expected for a rigid octahedron rotation. But the maximum of the probability stays at the high symmetry position ruling out an essential order-disorder character of the structural phase transition in K$_2$OsBr$_6$. The consistency of the anharmonic treatment was further verified by analyzing the residual electron densities. Significant amounts remain in the harmonic refinements close to the Br sites and are fully suppressed by the anharmonic treatment. 

Because of the significant anharmonicity of the Br ions, a refinement in the cubic phase has been attempted by considering a disordered Br site, i.e. a split model. In $Fm\bar{3}m$, the Br ions occupy the Wyckoff site $24e$ with atomic position ($x~0~0$). In the split model, an additional degree of freedom is introduced in order to characterize the disorder: the Br atomic positions are now ($x~y~0$) but the occupancy of the $24e$ Wyckoff site is maintained (or 1/4 of the full occupancy of the new Wyckoff site $96j$ with atomic position ($x~y~0$)). The refinement results are shown in Table~\ref{KOB_250K_order_disorder}. The refinement quality is only slightly improved compared to the one with the standard cubic model shown in Table~\ref{standard_refinement_KOB}. A significant non-zero deviation from the cubic symmetry is found $y=0.0158(9)$ but a strong correlation -- intrinsic to this kind of model -- occurs between the $y$(Br1) deviation and $U_{\perp}$(Br). The split displacement is still smaller than the root mean-square displacement in the perpendicular direction, which signifies that the maximum of the probability density stays at the high-symmetry value.

Both the moderate anharmonic ADPs yielding only slight changes in the probability distribution and
the failure of the split model to considerably improve the refinement quality exclude a predominant order-disorder character of the structural phase transition in K$_2$OsBr$_6$. The local character of the rotational modes in the antifluorite-type structure results in large and strongly anisotropic ADPs that remain however of dynamic origin.

\subsection{K$_2$IrBr$_6$}
\label{KIB_structure}

K$_2$IrBr$_6$ undergoes two structural transitions, first towards a tetragonal phase with SG $P4/mnc$ at $\approx 182$K, and then towards a monoclinic phase with space group $P2_1/n$ around $\approx 120$K where the monoclinic distortion is inferred from the temperature dependence of the lattice parameters, even though no macroscopic signature is visible~\cite{Reigiplessis20,Khan21}. The tetragonal distortion of the octahedra remains negligible at all temperatures while an orthorhombic distortion of $\approx 1$~\% develops in the monoclinic phase~\cite{Khan21}. Even though the rotary phonon mode describes the primary order parameter, the hysteretic behavior in thermal expansion measurement and the phase coexistence in powder XRD suggest that the upper structural transition possesses some 1st-order character driven by the emergence of a pronounced tetragonal strain, acting as a secondary order parameter~\cite{Khan21}. However this behavior does not imply an order-disorder scenario. 

Therefore, single-crystal XRD measurements have been performed at 250K and 200K in the cubic phase, see Table~\ref{Data_set_all} for an overview of the data set completeness. The lattice parameters at both temperatures are consistent with those from neutron and synchrotron powder diffraction measurements~\cite{Reigiplessis20,Khan21}. Refinements carried out in space group $Fm\bar{3}m$ with harmonic and anharmonic Br ADPs at both temperatures are reported in Table~\ref{KIB_refinements}. 
The transverse Br ADP value is similar to the one deduced for K$_2$OsBr$_6$. It is the strongest among the investigated compounds, probably due to the proximity of the structural phase transition in this temperature range for K$_2$IrBr$_6$. 
The anharmonic treatment improves the quality of the refinement, as shown by the pronounced reduction of the $R$-factors and the GOF parameters. 
Together with K$_2$OsBr$_6$, the anharmonic treatment in the cubic phase of K$_2$IrBr$_6$ has the most significant impact among the investigated compounds, see Table~\ref{anharmonic_param_all}. 
However, these parameters shrink upon cooling to 200K, when approaching the upper structural phase transition. In addition, the anharmonic treatment does not reduce the large $U_{\perp}$. Therefore, the anharmonic refinements do not support an order-disorder character.

RIXS measurements on K$_2$IrBr$_6$ evidenced a splitting of the spin-orbit exciton peak, a signature of the split low-lying $j=\frac{3}{2}$ quadruplet, and interpreted it as a fingerprint of a non-cubic crystal-field splitting $\Delta \approx 40-70$meV~\cite{Reigiplessis20,Khan21}. More importantly, this splitting already exists at room temperature, and the strength of $\Delta$ decreases with temperature while approaching the upper structural transition~\cite{Khan21}. These features were also confirmed by IR spectroscopy~\cite{Meggle23}. While the orthorhombic octahedra distortion appearing in the monoclinic phase, and the tetragonal strain in the intermediate phase can explain a non-cubic crystal-field splitting, the room-temperature feature was ascribed to the pronounced transverse motion of the Br ions breaking the cubic environment locally. However, at room temperature in the cubic phase, there was no evidence for a non-cubic crystal-field in K$_2$OsBr$_6$ in RIXS and IR spectroscopy~\cite{Warzanowski23}. Our XRD analysis shows that at 250K, the transverse motion of the Br ion deduced from the anharmonic refinement is $U_{\perp}=0.0621(4)$\AA$^2$ and $0.0585(2)$\AA$^2$ for K$_2$IrBr$_6$ and for K$_2$OsBr$_6$, respectively. Therefore, similar to K$_2$OsBr$_6$, the strong Br displacement perpendicular to the Ir-Br bond in the high temperature phase must be of dynamic origin. The local environment thus remains cubic and cannot explain the origin of a non-cubic distortion. Actually, recent works have reinterpreted the RIXS spectra of K$_2$RuCl$_6$~\cite{Iwahara23a} and K$_2$IrCl$_6$~\cite{Iwahara23b} in terms of the dynamic effects. 

\small
\begin{table}[h!]
	\centering
	\caption{\label{KIB_refinements} Summary of refinements in space group $Fm\bar{3}m$ with single-crystal XRD data measured at 250K and 200K on K$_2$IrBr$_6$. Refinement results with only harmonic ADPs, given in~\AA$^2$, are indicated on the left side of the arrow symbol ``$\rightarrow$", while the refinement including also anharmonic terms for the Br ions is reported on the right side. $R$-factors are given in \%.}
		\begin{indented}
	\lineup
	\item[]\begin{tabular}{@{}*{3}{l}}
		\br 
			&250K&200K \cr	
			\mr		
			$x$(Br)&0.23976(4)$\rightarrow$ 0.23940(4)&0.24014(4)$\rightarrow$0.23984(5)\cr
			$U_{\rm iso}$(Ir)&0.02427(6)$\rightarrow$ 0.02436(5)&0.01943(6)$\rightarrow$0.01946(6) \cr
			$U_{\rm iso}$(K)&0.0516(4)$\rightarrow$0.0513(3)&0.0418(4)$\rightarrow$0.0415(3) \cr
			$U_{\parallel}$(Br)&0.0242(1)$\rightarrow$0.0252(3)&0.0190(1)$\rightarrow$0.0202(3) \cr
			$U_{\perp}$(Br)&0.0613(2)$\rightarrow$0.0621(4)&0.0544(2)$\rightarrow$0.0541(4)	 \cr	
			\mr		
			$R(obs)$&2.12$\rightarrow$1.85& 2.48$\rightarrow$2.26 \cr
			$wR({\rm obs})$&2.32$\rightarrow$1.84&2.83$\rightarrow$2.43 \cr
			$R({\rm all})$&2.22$\rightarrow$1.94&2.48$\rightarrow$2.26 \cr
			$wR({\rm all})$&2.34$\rightarrow$1.86&2.83$\rightarrow$2.43 \cr
			GOF(obs)&1.50$\rightarrow$1.20&2.30$\rightarrow$1.99 \cr
			GOF(all)&1.48$\rightarrow$1.19&2.30 $\rightarrow$1.99 \cr
			\br  			
		\end{tabular}
\end{indented}		
\end{table}
\normalsize

\subsection{K$_2$OsCl$_6$}
\label{KOC_structure}

K$_2$OsCl$_6$ undergoes a structural phase transition at $T_c \approx 45~$K marked by a $\lambda$-type anomaly in specific heat measurements~\cite{Novotny77}. Inelastic neutron scattering experiments have evidenced the softening of an optical phonon branch, identified as a rotary phonon mode, at both the Brillouin zone center and at the Brillouin zone boundary (001). The $\Gamma$ mode is claimed to drive the structural phase transition since it softens more rapidly~\cite{Mintz79}. This is in line with neutron diffraction experiments revealing the emergence of the superstructure reflections (117) and (337) at the transition temperature which do not break the $F$ centering~\cite{Armstrong78}. Furthermore, no superstructure reflections breaking the $F$ centering have been found using neutron powder diffraction, and Rietveld refinements with these patterns were carried out with the tetragonal space group $I4/m$~\cite{SauraMuzquiz22}.

Single-crystal XRD measurements have been performed at both temperatures 250K and 80K in the cubic phase, see Table~\ref{Data_set_all} for an overview of the data set completeness. The room-temperature lattice parameter $a=9.7813(12)$\AA~agrees with powder XRD data~\cite{Fergusson74}. Refinements with data sets collected at both temperatures have been carried out in space group $Fm\bar{3}m$ and are summarized in Table~\ref{KOC_refinements}. $R$-factors and goodness-of-fit (GOF) parameters indicate the good quality of the refinements. We found that $U_{\perp} \gg U_{\parallel}$, in line with the impact of soft rotation modes. A refinement with anharmonic parameters up to rank four for the Br ions has been carried out.
At 250K, the anharmonic refinement does not clearly improve the $R$ values of the harmonic refinement, and none of the refined anharmonic parameters is significant, see Table~\ref{anharmonic_param_all}. 
However, when approaching the transition temperature at 80\,K, the GOF parameters and the $wR$-values are slightly improved compared to the harmonic refinement. The amplitude of the two Br harmonic ADPs is reduced, and two anharmonic ADPs --- $D_{1111}$ and $D_{2222}=D_{3333}$ --- exceed $3\sigma$, see Table~\ref{anharmonic_param_all}. 
In agreement with the softening of the phonon rotary mode at the Brillouin zone center pointing to a displacive structural phase transition, anharmonic effects remain moderate in the high-temperature phase of  K$_2$OsCl$_6$. 
In view of possible local distortions as an explanation for the large transversal ADP of the Cl, one has to inspect its temperature dependence. 
The very large $U_{\perp}$ value that at 250K corresponds to a root mean-square displacement of more than 0.2 \AA ~ strongly shrinks upon cooling to 80K excluding a local disorder scenario. 
\small
\begin{table}[t!]
	\centering
	\caption{\label{KOC_refinements} Summary of refinements in space group $Fm\bar{3}m$ with single-crystal XRD data measured at 250K and 80K on K$_2$OsCl$_6$. Refinement results with only harmonic ADPs, given in~\AA$^2$, are indicated on the left side of the arrow symbol ``$\rightarrow$", while refinement including also anharmonic terms for the Br ions are reported on the right side. $R$-factors are given in \%.}
		\begin{indented}
	\lineup
	\item[]\begin{tabular}{@{}*{3}{l}}
		\br  
		&250K&80K \cr	
		\mr		
		$x$(Cl)&0.2385(1)$\rightarrow$0.2382(2)&0.2403(1)$\rightarrow$0.2405(2)\cr
		$U_{\rm iso}$(Os)&0.01685(9)$\rightarrow$0.01685(9)&0.0073(1)$\rightarrow$0.00721(9) \cr
		$U_{\rm iso}$(K)&0.0368(4)$\rightarrow$0.0368(4)& 0.0150(3)$\rightarrow$0.0151(2) \cr
		$U_{\parallel}$(Cl)&0.0174(4)$\rightarrow$0.0174(4)&0.0085(3)$\rightarrow$0.0053(8) \cr
		$U_{\perp}$(Cl)&0.0421(4)$\rightarrow$0.0416(5)&0.0191(3)$\rightarrow$0.0171(6)	 \cr	
		\mr		
		$R({\rm obs})$&2.18$\rightarrow$2.15& 1.96$\rightarrow$1.99 \cr
		$wR({\rm obs})$&2.88$\rightarrow$2.84&3.06$\rightarrow$2.80 \cr
		$R({\rm all})$&2.18$\rightarrow$2.15&1.96$\rightarrow$1.99 \cr
		$wR({\rm all})$&2.88$\rightarrow$2.84&3.06$\rightarrow$2.80 \cr
		GOF(obs)&2.54$\rightarrow$2.52&2.80$\rightarrow$2.58 \cr
		GOF(all)&2.54$\rightarrow$2.52&2.80$\rightarrow$2.58 \cr
		\br  			
	\end{tabular}
\end{indented}	
\end{table}
\normalsize

\subsection{K$_2$IrCl$_6$}
\label{KIC_structure}

Synchrotron powder XRD measurements have shown that K$_2$IrCl$_6$ remains cubic down to 20K, and no evidence in thermodynamic measurements point to a structural phase transition down to the N\'eel temperature $T_{\rm N}=3$K~\cite{Khan19}. Later on, neutron powder diffraction experiments indicated that the structure remains cubic while crossing $T_{\rm N}$~\cite{Reigiplessis20}. High-field ESR measurements result in a $g$ factor consistent with a local cubic environment in the paramagnetic state~\cite{Bhaskaran21}. Despite the absence of structural phase transitions, the splitting of the low lying $j=\frac{3}{2}$ quartet into at least two doublets already at room temperature as observed with RIXS spectroscopy~\cite{Reigiplessis20}, IR spectroscopy~\cite{Meggle23}, and Raman spectroscopy~\cite{Lee22}, is interpreted as a deviation from the local cubic environment at the Ir site, with a non cubic crystal-field $\Delta \approx 50-60$~meV slightly stronger than in the bromide counterpart~\cite{Reigiplessis20}. 

In order to investigate the local symmetry breaking, single-crystal XRD measurements have been performed at 200K and 80K, see Table~\ref{Data_set_all} for an overview of the data set completeness. The lattice parameters at both temperatures are consistent with those reported from high-resolution synchrotron XRD~\cite{Khan19}. Refinements carried out in space group $Fm\bar{3}m$ with harmonic and anharmonic ADPs at both temperatures are reported in Table~\ref{KIC_refinements}. First, if $U_{\perp}$(Cl) is consistent with the value reported in literature~\cite{Khan19}, $U_{\parallel}$(Cl) is larger at 200K and shrinks upon cooling compared to the temperature independent behavior reported in~\cite{Khan19}. Also, the Cl ADPs are slightly smaller than the Br ones in K$_2$OsBr$_6$ at 80K. Even though Cl atoms are lighter elements, a displacive structural transition is absent in the Ir case. The anharmonic treatment does not improve the quality of the refinement at 200K, and very little at 80K. Note however that at 80K, the anharmonic treatment induces a reduction of the harmonic Cl ADPs, as it is also seen in K$_2$OsCl$_6$. This observation was not found in any other of the investigated compounds. The anharmonic parameters listed in Table~\ref{anharmonic_param_all} are much weaker compared to the Br counterpart. Therefore, the relevance or not of anharmonic effects in the atomic displacements is a reliable indicator of a subsequent/absent structural phase transition at lower temperature. 
 
Following the conclusions of Sec.~\ref{KIB_structure}, there is no indication of local disorder in K$_2$IrCl$_6$, and the splitting observed in the RIXS spectra cannot be accounted for by the static structure. Instead dynamic processes should be taken into consideration.

\small
\begin{table}[h!]
	\centering
	\caption{\label{KIC_refinements} Summary of refinements in space group $Fm\bar{3}m$ with single-crystal XRD data measured at 200K and 80K on K$_2$IrCl$_6$. Refinement results with only harmonic ADPs, given in~\AA$^2$, are indicated on the left side of the arrow symbol ``$\rightarrow$", while the refinement including also anharmonic terms for the Br ions is reported on the right side. $R$-factors are given in \%.}
		\begin{indented}
	\lineup
	\item[]\begin{tabular}{@{}*{3}{l}}
		\br 
			&200K&80K \cr	
			\mr		
			$x$(Cl)&0.23867(7)$\rightarrow$0.23833(9)&0.23992(7)$\rightarrow$0.23992(7)\cr
			$U_{\rm iso}$(Ir)&0.01140(4)$\rightarrow$0.01140(4)&0.00431(4)$\rightarrow$0.00427(4) \cr
			$U_{\rm iso}$(K)&0.0273(2)$\rightarrow$0.0273(2)& 0.0118(1)$\rightarrow$0.0118(1) \cr
			$U_{\parallel}$(Cl)&0.0125(2)$\rightarrow$0.0125(2)&0.0057(2)$\rightarrow$0.0042(5) \cr
			$U_{\perp}$(Cl)&0.0308(2)$\rightarrow$0.0308(2)&0.0144(1)$\rightarrow$0.0130(3)	 \cr	
			\mr		
			$R({\rm obs})$& 1.17$\rightarrow$1.12& 1.71$\rightarrow$1.69 \cr
			$wR({\rm obs})$& 1.78$\rightarrow$1.72&2.13$\rightarrow$1.99 \cr
			$R({\rm all})$& 1.17$\rightarrow$1.12&1.71$\rightarrow$1.69 \cr
			$wR({\rm all})$& 1.78$\rightarrow$1.72&2.13$\rightarrow$1.99 \cr
			GOF(obs)&1.40$\rightarrow$1.35&1.85$\rightarrow$1.74 \cr
			GOF(all)&1.40$\rightarrow$1.35&1.85$\rightarrow$1.74 \cr
			\br 			
		\end{tabular}
\end{indented}		
\end{table}
\normalsize


\subsection{Rb$_2$OsBr$_6$}
\label{ROB_structure}

\par
Single-crystal XRD measurements have been performed at both temperatures 250K and 80K, see Table~\ref{Data_set_all} for an overview of the data set completeness. At room temperature, Rb$_2$OsBr$_6$ crystallizes in the $fcc$ cubic structure $Fm\bar{3}m$. The room-temperature lattice parameter $a=10.4265(15)$\AA~agrees with powder XRD data~\cite{Fergusson74}. A structural sequence of phase transitions is not documented for Rb$_2$OsBr$_6$, and temperature dependent powder diffraction data are lacking. However, no hints for structural phase transitions are visible in our experiments since no superstructure reflections could be resolved in the computed precession maps upon cooling down to 80K. Thus, refinements at both temperatures have been carried out with space group $Fm\bar{3}m$ and are summarized in Table~\ref{ROB_refinements}. $R$-factors and GOF parameters indicate the good quality of the refinements, and confirmed the cubic $Fm\bar{3}m$ structure down to at least 80K. Again, we found that $U_{\perp} \gg U_{\parallel}$. As in the two other cases, the relative large amplitude of the Br ADP perpendicular to the bond stems from the softness of the rotation modes. Therefore, a refinement with anharmonic ADPs up to rank four for the Br ions has been carried out. Already at 250K, and without the proximity of a structural phase transition, $R$-values of the anharmonic refinement are slightly improved compared to the refinement with only harmonic ADPs, see Table~\ref{ROB_refinements}. The refined anharmonic parameters are tabulated in Table~\ref{anharmonic_param_all}. 
Only $D_{2222}=D_{3333}$, probing the anharmonicity perpendicular to the Os-Br bond, attains a significant value compared to its standard deviation. However, $D_{2222}=D_{3333}$ decreases with lower temperatures and therefore does not agree with the premises of an order-disorder structural phase transition at low temperature. Also in this material the anomalous transversal ADP
$U_{\perp}$ of the halide strongly shrinks upon cooling indicating its fully dynamic origin.
\small
\begin{table}[h!]
	\centering
	\caption{\label{ROB_refinements} Summary of refinements in space group $Fm\bar{3}m$ with single-crystal XRD data measured at 250K and 80K on Rb$_2$OsBr$_6$. Refinement results with only harmonic ADPs, given in~\AA$^2$, are indicated on the left side of the arrow symbol ``$\rightarrow$", while the refinement including also anharmonic terms for the Br ions is reported on the right side. $R$-factors are given in \%.}
		\begin{indented}
	\lineup
	\item[]\begin{tabular}{@{}*{3}{l}}
		\br  
			&250K&80K \cr	
			\mr		
			$x$(Br)&0.23882(4)$\rightarrow$0.23862(5)&0.24032(4)$\rightarrow$0.24017(5)\cr
			$U_{\rm iso}$(Os)&0.0213(1)$\rightarrow$0.02142(9)&0.01003(8)$\rightarrow$0.01014(8) \cr
			$U_{\rm iso}$(Rb)&0.0372(2)$\rightarrow$0.0374(2)& 0.0156(1)$\rightarrow$0.0158(1) \cr
			$U_{\parallel}$(Br)&0.0216(2)$\rightarrow$0.0219(2)&0.0104(1)$\rightarrow$0.0109(4) \cr
			$U_{\perp}$(Br)&0.0373(2)$\rightarrow$0.0400(4)&0.0159(1)$\rightarrow$0.0174(3)	 \cr	
			\mr		
			$R({\rm obs})$& 1.76$\rightarrow$1.60& 1.94$\rightarrow$1.82 \cr
			$wR({\rm obs})$& 1.78$\rightarrow$1.50&2.56$\rightarrow$2.40 \cr
			$R({\rm all})$& 1.79$\rightarrow$1.63&2.09$\rightarrow$1.98 \cr
			$wR({\rm all})$& 1.79$\rightarrow$1.51&2.59$\rightarrow$2.43 \cr
			GOF(obs)&1.49$\rightarrow$1.27&2.08$\rightarrow$1.97 \cr
			GOF(all)&1.46$\rightarrow$1.25&2.06$\rightarrow$1.95 \cr
			\br  			
		\end{tabular}
\end{indented}		
\end{table}
\normalsize

\subsection{Cs$_2$OsBr$_6$}
\label{COB_structure}
Except for IR spectroscopy measurements at 2K~\cite{Schmidtke97} stressing the resemblance of absorption spectra with those reported for Os$^{4+}$ doped in Cs$_2$ZrBr$_6$ which preserve the octahedral environment at the Os site~\cite{Kozinowski80}, the crystal structure of Cs$_2$OsBr$_6$ is not documented. Single-crystal XRD measurements have been performed at both temperatures 250K and 80K, see Table~\ref{Data_set_all} for an overview of the data set completeness. At 250K, the crystal structure is cubic, and no superstructure reflections are visible down to 80K. Therefore refinements at both temperatures are performed with space group $Fm\bar{3}m$. Structural parameters are summarized in Table~\ref{COB_refinements}, together with $R$-factors and GOF parameters indicative of the good refinement quality and confirming the correct space group assignment. The dynamical behavior of the bromide ion is very similar to Rb$_2$OsBr$_6$, see Section~\ref{ROB_structure}. Analogous refinements with anharmonic ADPS up to rank four for the Br ion are carried out, but none of them attains a significant values, see Table~\ref{anharmonic_param_all}. Although $U_{\perp} \gg U_{\parallel}$ at 250K, $U_{\perp}$ considerably shrinks upon cooling and the Br thermal ellipsoid becomes almost isotropic. Therefore, there are no hints for a low temperature structural distortion of neither long-range nor local character.
\small
\begin{table}[h!]
	\centering
	\caption{\label{COB_refinements} Summary of refinements in space group $Fm\bar{3}m$ with single-crystal XRD data measured at 250K and 80K on Cs$_2$OsBr$_6$. Refinement results with only harmonic ADPs, given in~\AA$^2$, are indicated on the left side of the arrow symbol ``$\rightarrow$", while the refinement including also anharmonic terms for the Br ions is reported on the right side. $R$-factors are given in \%.}
		\begin{indented}
	\lineup
	\item[]
	\begin{tabular}{@{}*{3}{l}}
		\br 
			&250K&80K \cr	
			\mr		
			$x$(Br)&0.23464(6)$\rightarrow$0.2350(1)&0.23618(7)$\rightarrow$0.2364(1)\cr
			$U_{\rm iso}$(Os)&0.0195(2)$\rightarrow$0.0196(2)&0.0092(2)$\rightarrow$0.0094(2) \cr
			$U_{\rm iso}$(Cs)&0.0303(2)$\rightarrow$0.0303(2)& 0.0140(2)$\rightarrow$0.0141(2) \cr
			$U_{\parallel}$(Br)&0.0212(3)$\rightarrow$0.0225(7)&0.0110(3)$\rightarrow$0.0126(9) \cr
			$U_{\perp}$(Br)&0.0310(2)$\rightarrow$0.0306(3)&0.0141(2)$\rightarrow$0.0149(6)	 \cr	
			\mr		
			$R({\rm obs})$& 1.32$\rightarrow$1.23& 1.34$\rightarrow$1.26 \cr
			$wR({\rm obs})$& 1.70$\rightarrow$1.55&1.62$\rightarrow$1.55 \cr
			$R({\rm all})$& 1.41$\rightarrow$1.32&1.59$\rightarrow$1.50 \cr
			$wR({\rm all})$& 1.75$\rightarrow$1.61&1.68$\rightarrow$1.61 \cr
			GOF(obs)&1.32$\rightarrow$1.23&1.14$\rightarrow$1.11 \cr
			GOF(all)&1.33$\rightarrow$1.25&1.11$\rightarrow$1.09 \cr
			\br  			
		\end{tabular}
\end{indented}		
\end{table}
\normalsize

\subsection{(NH$_4$)$_2$OsBr$_6$}
\label{NHOB_structure}

The replacement of the cation $A$ by an ammonium group NH$_4$ introduces additional degrees of freedom. The rotational orientation of the NH$_4$ group can
give rise to structural transitions with order-disorder character. The tendency towards an ordered state is enhanced in deuterated samples simply due to the larger mass. For instance, in (NH$_4$)$_2$TeCl$_6$, the deuterated sample exhibits two additional structural transitions interpreted as localization of the deuterium in a triangular potential well, while in the protonated counterpart the tetrahedra remain fluctuating~\cite{Dimitropoulos90,Kume91,Arnscheidt92}. An other striking example is the order-disorder transition in (ND$_4$)$_2$PdCl$_6$, which is absent in (NH$_4$)$_2$PdCl$_6$~\cite{Swainson97}.

(NH$_4$)$_2$OsBr$_6$ crystallizes at room temperature in the common antifluorite-type $Fm\bar{3}m$ structure~\cite{Fergusson74,Kohlmann22}. No hint for a structural transition is visible in differential scanning calorimetry and thermogravimetry from 113K up to the decomposition temperature at 673K, and powder XRD data confirm the cubic structure with  $Fm\bar{3}m$ from room temperature up to 673K~\cite{Kohlmann22}.

\begin{figure}[ht!]
		\includegraphics[width=0.49\columnwidth]{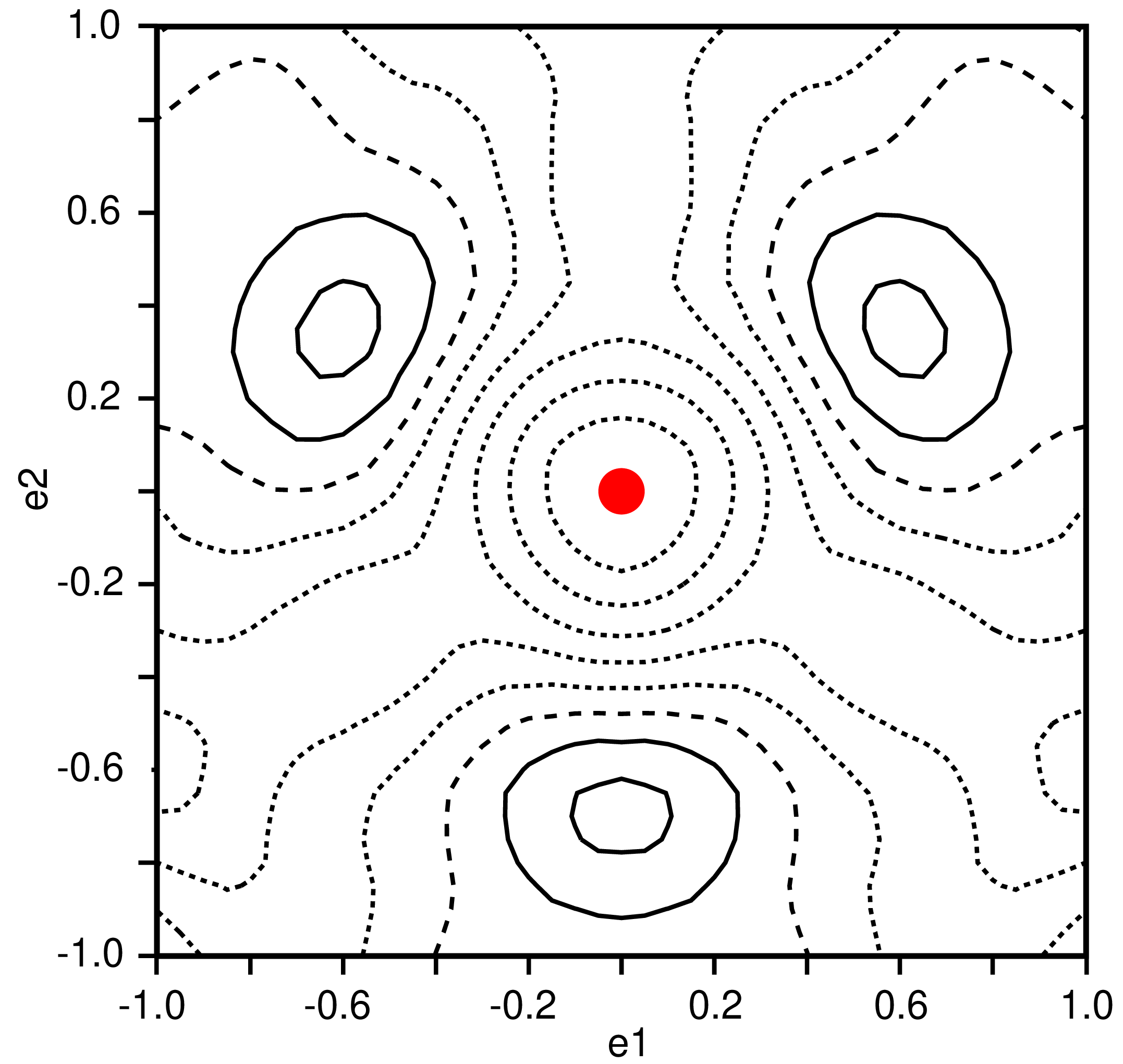}		
		\includegraphics[width=0.49\columnwidth]{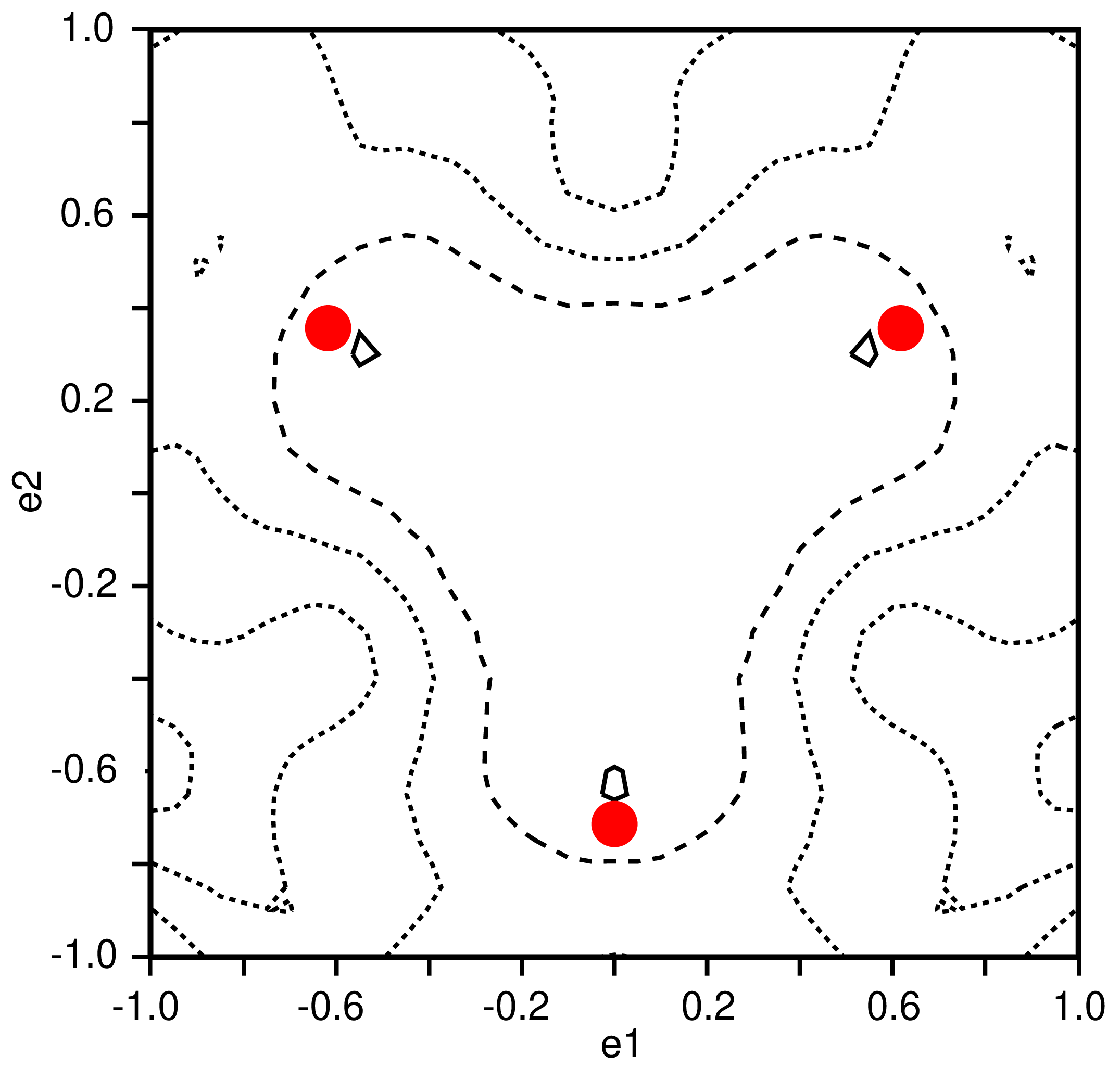}
	\makeatletter\long\def\@ifdim#1#2#3{#2}\makeatother
	\caption{\label{NHOB_maps} Residual scattering density at 200K computed in the (111) plane, defined with {\bf e1} and {\bf e2} two orthogonal vectors in real space (\AA), and with the center of the map being the $32f$ Wyckoff position with $x_{\rm H}=0.208$. In red are indicated the H positions from the two different models. The dotted and full lines illustrate the negative and positive RSD, respectively. The contour step scaling is the same in both panels. Left: residual density computed with the {\sl normal} tetrahedra model M1 where the H atom is sitting at $x_{\rm H}=0.208$, close to the central Os ion. Right: computed with the {\sl inverted} tetrahedra model M2 ($x_{\rm H}=0.292$).}
\end{figure}

Single-crystal XRD measurements have been performed at 200K and 80K, see Table~\ref{Data_set_all} for an overview of the data set completeness. No superstructure reflections were visible at both temperatures and refinements were performed with the cubic space group $Fm\bar{3}m$. 
Because of the very weak X-ray form factor of the H atom, the isotropic Debye-Waller factor used in our refinements was linearly scaled from the value $U_{\rm iso}=0.025$~\AA$^2$ deduced from powder XRD at room temperature~\cite{Kohlmann22}, i.e.\ $U_{\rm iso}=0.017$~\AA$^2$ and $U_{\rm iso}=0.007$~\AA$^2$ at 200K and 80K, respectively. Prior to focusing on the dynamics of the OsBr$_6$ octahedra, the static ordering of the NH$_4$ tetrahedra will be discussed. In the ammonium antifluorites, the tetrahedra are oriented in such a way that the N-H bonds are aligned along the 3-fold axis [111]. Following the methodology of~\cite{Yamamuro97}, two ordered models are considered:
\begin{enumerate}
	\item[.] 
In the {\sl normal} tetrahedron orientation (model M1) the H atoms point towards the nearest neighboring Os ions 
	\item[.]
In the {\sl inverted} tetrahedron orientation  (model M2), H atoms are closer to the empty $B'$ sites of the double perovskite structure, as it is illustrated in Fig~\ref{structure_fig}(b)	
\end{enumerate}
Note that model M2 results from model M1 by applying a space inversion with respect to the central N position. Refinement results in $Fm\bar{3}m$ with these two models at both temperatures are summarized in Table~\ref{split_NHOB}. The main difference between these models is the position of the H atom occupying the $32f$ Wyckoff site: $x_{\rm H}=0.208(2)$ and 0.292(1) at 200K for model M1 and model M2, respectively. Regarding the $R$-factors and the GOF parameters, it is clear that model M2 best describes the data set. For completeness, the residual scattering density at 200K, calculated as $F^2$({\rm obs})-$F^2$({\rm calc}), is computed in the (111) plane with the center being the $32f$ Wyckoff position ($x_{\rm H}=0.208$) for model M1 (left panel) and model M2 (right panel) in Figure~\ref{NHOB_maps}. For model M1, there is a maximum of negative RSD at the center, where the H atom shown in red is located, surrounded by three maxima of positive RSD. When applying model M2  the map does not show any residual scattering density anymore~\footnote{At 200K, a superposition of the two static models give an H occupation of 34(11)\% and 66(11)\% for model M1 and M2, respectively. Model M2 is still favored but the H occupation for model M1 does not fully vanish.}. This inverted ordering of the tetrahedra was found for instance in (NH$_4$)$_2$TeCl$_6$~\cite{Armstrong91}. Powder XRD measurements concluded that (NH$_4$)$_2$OsBr$_6$ adopts the fully ordered ammonium molecule corresponding to model M1 in our discussion~\cite{Kohlmann22}. However, the two different tetrahedra orientations have not been investigated. The refinements presented in Section~\ref{sec_crystal} checking the stoichiometry of our sample, but also  forthcoming refinements with anharmonic ADPs for Br ions, are performed starting with H positional parameters of model M2.
\small
\begin{table*}[ht!]
	\centering
	\caption{\label{split_NHOB}  Summary of refinements in space group $Fm\bar{3}m$ with single-crystal XRD data measured at 200K and 80K on (NH$_4$)$_2$OsBr$_6$. Two models M1 and M2 are considered in order to investigate the tetrahedra static ordering, see main text. On the left and right side of the symbol ``/" we report the $R$-factors, given in \%, and GOF parameters for the observed ($I/\sigma > 3$) and for all reflections, respectively. ADPs are given in~\AA$^2$.}
		\begin{indented}
	\lineup
	\item[]
	\begin{center}
	\begin{tabular}{@{}*{8}{l}}
		\br  
		&&\multicolumn{3}{c}{200K}&\multicolumn{3}{c}{80K} \cr	
		\mr
		\multirow{4}{*}{M1}&$x_{\rm H}~x_{\rm Br}~U_{\rm iso}^{\rm N}$ &0.208(1)&0.24005(2)&0.0273(6)&0.208(1)&0.24112(2)&0.0127(3)	 \cr
		&$U_{\parallel}^{\rm Br}~U_{\perp}^{\rm Br}~U_{\rm iso}^{\rm Os}$&0.01289(6)&0.02795(6)&0.01299(4)&0.00619(5)&0.01406(5)&0.00603(3) \cr
		& $R~wR~$GOF(obs/all)& 1.21/1.33& 1.55/1.58&1.15/1.14&0.99/1.08 & 1.53/1.56&1.27/1.26 \cr
		\mr  	
		\multirow{4}{*}{M2}&$x_{\rm H}~x_{\rm Br}~U_{\rm iso}^{\rm N}$ &0.292(1)&0.24000(2)&0.0274(5)&0.292(1)&0.24108(1)&0.0127(3)	 \cr
		&$U_{\parallel}^{\rm Br}~U_{\perp}^{\rm Br}~U_{\rm iso}^{\rm Os}$&0.01286(6)&0.02793(6)&0.01298(3)&0.00617(4)&0.01405(4)&0.00603(3)\cr
		& $R~wR~$GOF(obs/all)&1.16/1.28&1.42/1.45&1.05/1.05&0.92/1.01&1.35/1.37&1.11/1.12 \cr
		\br  	
	\end{tabular}
\end{center}	
\end{indented}	
\end{table*}
\normalsize

In a last step, we investigate the dynamics of the OsBr$_6$ octahedra. Starting from model M2, refinements with harmonic and anharmonic Br ADPs  at both temperatures are compared in Table~\ref{NHOB_refinements}. Contrary to the two other discussed Os compounds showing no indication of structural transitions --- Cs$_2$OsBr$_6$ and Rb$_2$OsBr$_6$ --- where the Br ADPs become almost isotropic upon cooling, in (NH$_4$)$_2$OsBr$_6$ $U_{\perp} \approx 2 \times U_{\parallel}$ at 80K, a situation reminiscent of K$_2$OsCl$_6$, which shows a structural transition at lower temperature. Anharmonic treatment of Br ADPs at 200K slightly improve the refinement quality, and the relevant anharmonic parameters are gathered in Table~\ref{anharmonic_param_all}. However, contrary to  K$_2$OsCl$_6$, there is no improvement of the harmonic refinement with anharmonic parameters upon cooling and the anharmonic parameters vanish at 80K.
Similar to the discussion above, this excludes local distortions playing a significant role. 

\small
\begin{table}[h!]
	\centering
	\caption{\label{NHOB_refinements} Summary of refinements in space group $Fm\bar{3}m$ with single-crystal XRD data measured at 200K and 80K on (NH$_4$)$_2$OsBr$_6$. Tetrahedra are oriented according to model M2, see main text. Refinement results with only harmonic ADPs, given in~\AA$^2$, are indicated on the left side of the arrow symbol ``$\rightarrow$", while the refinement including also anharmonic terms for the Br ions is reported on the right side. $R$-factors are given in \%.}
		\begin{indented}
	\lineup
	\item[]\begin{tabular}{@{}*{3}{l}}
		\br 
			&200K&80K \cr	
			\mr		
			$x$(Br)&0.23400(2)$\rightarrow$0.23987(2)&0.24108(1)$\rightarrow$0.24108(1)\cr
			$x$(H)&0.292(1)$\rightarrow$0.294(1)&0.292(1)$\rightarrow$0.293(1)\cr			
			$U_{\rm iso}$(Os)&0.01298(3)$\rightarrow$0.01303(3)&0.00603(3)$\rightarrow$0.00604(3) \cr
			$U_{\rm iso}$(N)&0.0274(5)$\rightarrow$0.0271(5)& 0.0127(3)$\rightarrow$0.0127(3) \cr
			$U_{\parallel}$(Br)&0.01286(6)$\rightarrow$0.01280(5)&0.00617(4)$\rightarrow$0.00614(5) \cr
			$U_{\perp}$(Br)&0.02793(6)$\rightarrow$0.0284(1)&0.01405(4)$\rightarrow$0.01423(8)	 \cr	
			\mr		
			$R({\rm obs})$&1.16$\rightarrow$1.08& 0.92$\rightarrow$0.91 \cr
			$wR({\rm obs})$&1.42$\rightarrow$1.28&1.35$\rightarrow$1.33 \cr
			$R({\rm all})$&1.28$\rightarrow$1.19&1.01$\rightarrow$1.01 \cr
			$wR({\rm all})$&1.45$\rightarrow$1.31&1.37$\rightarrow$1.36 \cr
			GOF(obs)&1.05$\rightarrow$0.95&1.11$\rightarrow$1.10 \cr
			GOF(all)&1.05$\rightarrow$0.95&1.12$\rightarrow$1.11 \cr
			\br			
		\end{tabular}
\end{indented}		
\end{table}
\normalsize

\section{Discussion}

The tendency of antifluorite-type compounds to distort has been initially discussed in accordance with the radius ratio, defined as the ionic radii of the cation $A$ divided by the cavity space available for $A$~\cite{Brown64}. The smaller the radius ratio is, the more distorted are the antifluorite-type compounds at room temperature. Authors of~\cite{Roessler77} pointed out that the structural transition temperature decreases with decreasing the anion $X$ size, varying the metal ion $Me$ across the $4d$/$5d$ shells, and increasing the cation $A$ size. A more quantitative analysis is obtained in terms of the ionic radii $r_i$ of the different ions and the perovskite Goldschmidt's tolerance factor ($t=\frac{r_X+r_A}{\sqrt{2}(r_M+r_X)}$) that describes a well-known criterion for the occurrence of rotational phase transitions in $ABX_3$ perovskites~\cite{Cai17,Fedorovskiy2020,Rahim20}.

Motivated by these previous studies, we discuss the structural instability in the antifluorite-type compounds in terms of the tolerance factor $t$ by analyzing the octahedra rotation and tilting angles and the large ADP  $U_{\perp}$($X$) perpendicular to the $Me-X$ bond. Additionally to the data obtained in this work, we include our single-crystal XRD data for K$_2$ReCl$_6$ and K$_2$SnCl$_6$~\cite{Bertin23}. 
We, furthermore, include data from the literature as it is specified in the note \footnote{Structural information reported in the main text: K$_2$WCl$_6$~\cite{Xu05}; K$_2$SeBr$_6$~\cite{Noda80,Abriel90}; K$_2$PtI$_6$~\cite{Bennet24}; K$_2$PtBr$_6$~\cite{Grundy69,Armstrong85}; K$_2$PtCl$_6$~\cite{Schefer98}; K$_2$ReBr$_6$~\cite{Grundy69}; K$_2$SnBr$_6$~\cite{Higashi79}; K$_2$TeBr$_6$~\cite{Brown64}; K$_2$TeI$_6$~\cite{Syoyama72, Peresh05}; K$_2$TcCl$_6$~\cite{Elder67}; K$_2$TaCl$_6$~\cite{Ishikawa19}; K$_2$ReI$_6$~\cite{Gui19}; K$_2$NbCl$_6$~\cite{Henke07}; (NH$_4$)$_2$PtI$_6$~\cite{Sutton81,Mcelroy80}; (NH$_4$)$_2$PtBr$_6$\cite{Sutton81,Weir02,Mcelroy78}; (NH$_4$)$_2$RuCl$_6$~\cite{Weir02};(NH$_4$)$_2$TeCl$_6$~\cite{Armstrong90,Hazell66,Kawald88} (transition to a trigonal space group, octahedra rotation angle around (111)$_c$); (NH$_4$)$_2$TeBr$_6$~\cite{Das66,Das69};  (NH$_4$)$_2$TeI$_6$~\cite{Abriel89,Furukawa89}; (NH$_4$)$_2$ReI$_6$~\cite{Gonzalez03,Kochel04}; (NH$_4$)$_2$SnBr$_6$\cite{Negita80,Negita82,Armstrong83}; (NH$_4$)$_2$SnBr$_6$~\cite{Brill74,Lerbscher76}; (NH$_4$)$_2$PbCl$_6$~\cite{Armstrong89b} (transition to a trigonal space group, octahedra rotation angle around (111)$_c$); Rb$_2$PtI$_6$~\cite{Sutton81,Mcelroy78}; Rb$_2$TeBr$_6$~\cite{Abriel84}; Rb$_2$TeI$_6$~\cite{Abriel82}; Rb$_2$TaCl$_6$~\cite{Ishikawa19}; Rb$_2$NbCl$_6$~\cite{Henke07} ($U_{\perp}$ at 120K, scaled at 250K); Cs$_2$PtI$_6$~\cite{Bennet24}; Cs$_2$TaCl$_6$~\cite{Ishikawa19,Yun07} ($U_{\perp}$ at 200K, scaled at 250K); Cs$_2$NbCl$_6$~\cite{Henke07}. If not documented elsewhere, transition temperature $T_S$ are taken from~\cite{Roessler77}.}. 
We analyze three entities: the structural transition temperature $T_S$ from the high temperature cubic phase to a lower symmetry phase, the anisotropic ADP $U_{\perp}$ taken from available room-temperature XRD or from neutron diffraction measurements and linearly scaled to $T=250$K, and the octahedra rotation angle $\phi$ around the $c$ axis and the tilt angle $\theta$ around the monoclinic unique axis $b_m \equiv (110)_c$ taken at the lowest available temperature. In the case of a symmetry lower than tetragonal, $\phi$ is the average rotation angle calculated from the octahedra basal plane ions $X_1$ and $X_2$, and $\theta=\theta_{\rm apical}$ is calculated with the apical $X_3$ ion.

\begin{figure}[ht!]
	\includegraphics[width=0.99\columnwidth]{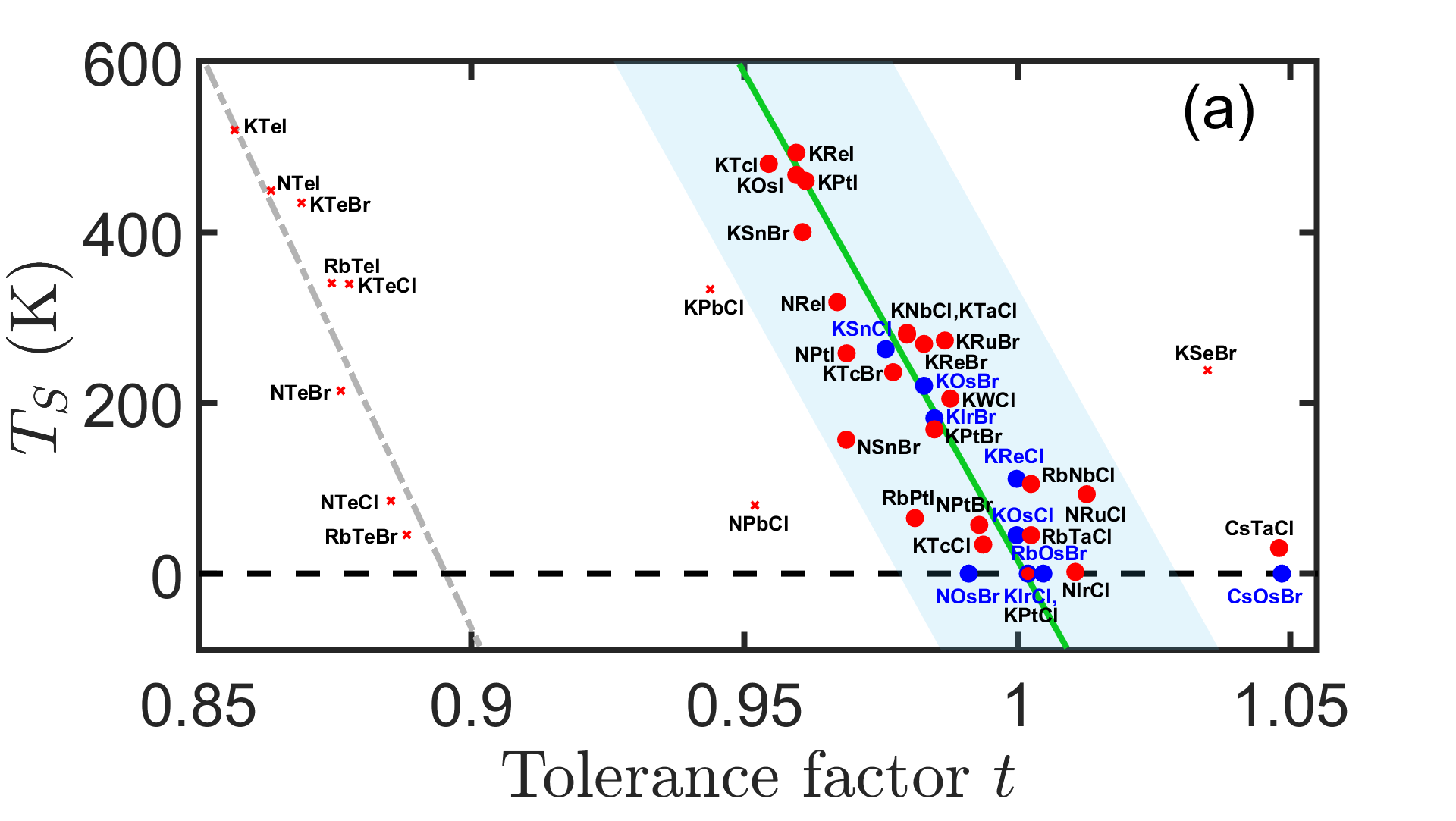}
	\includegraphics[width=0.99\columnwidth]{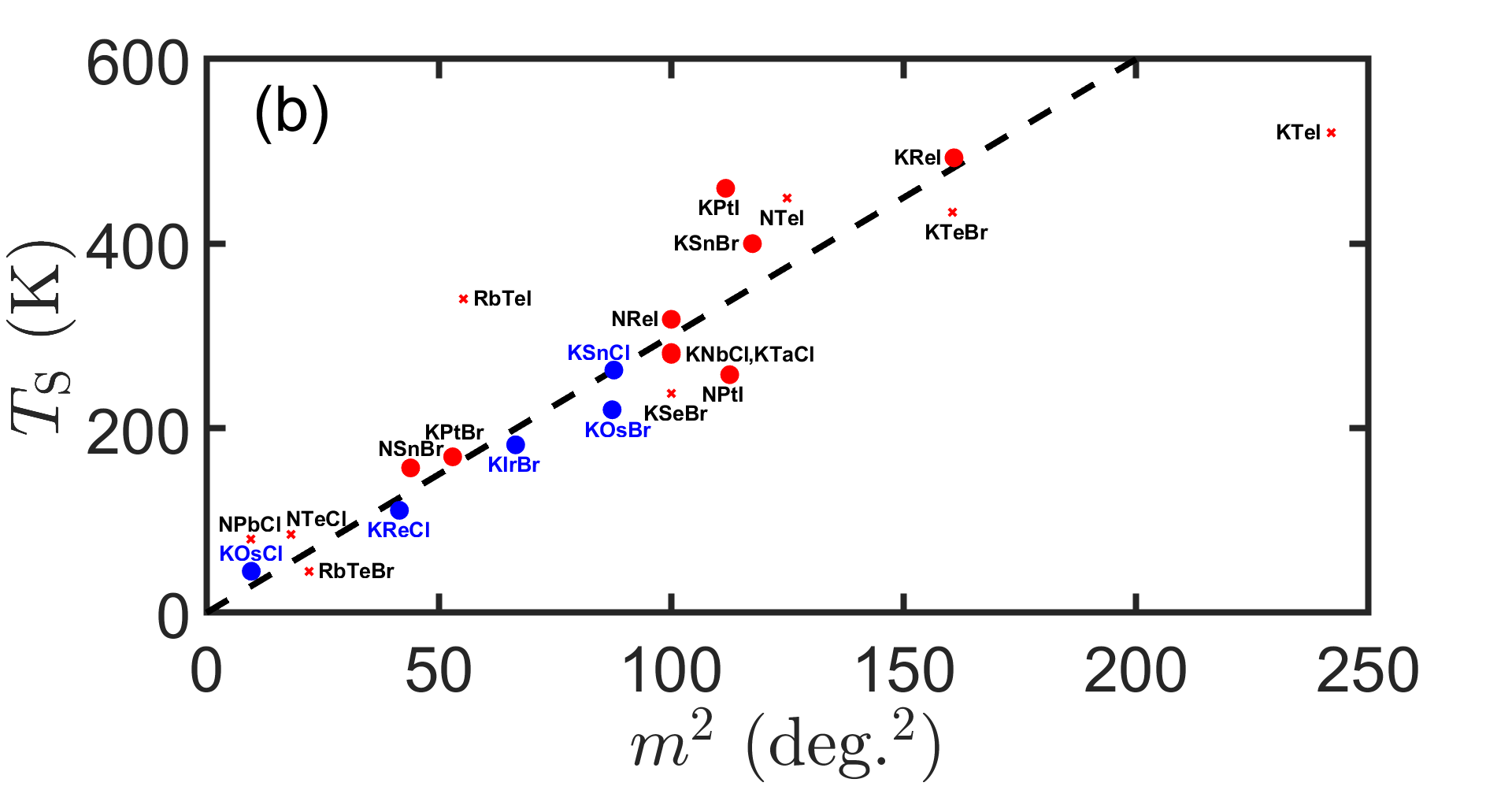}
	\includegraphics[width=0.99\columnwidth]{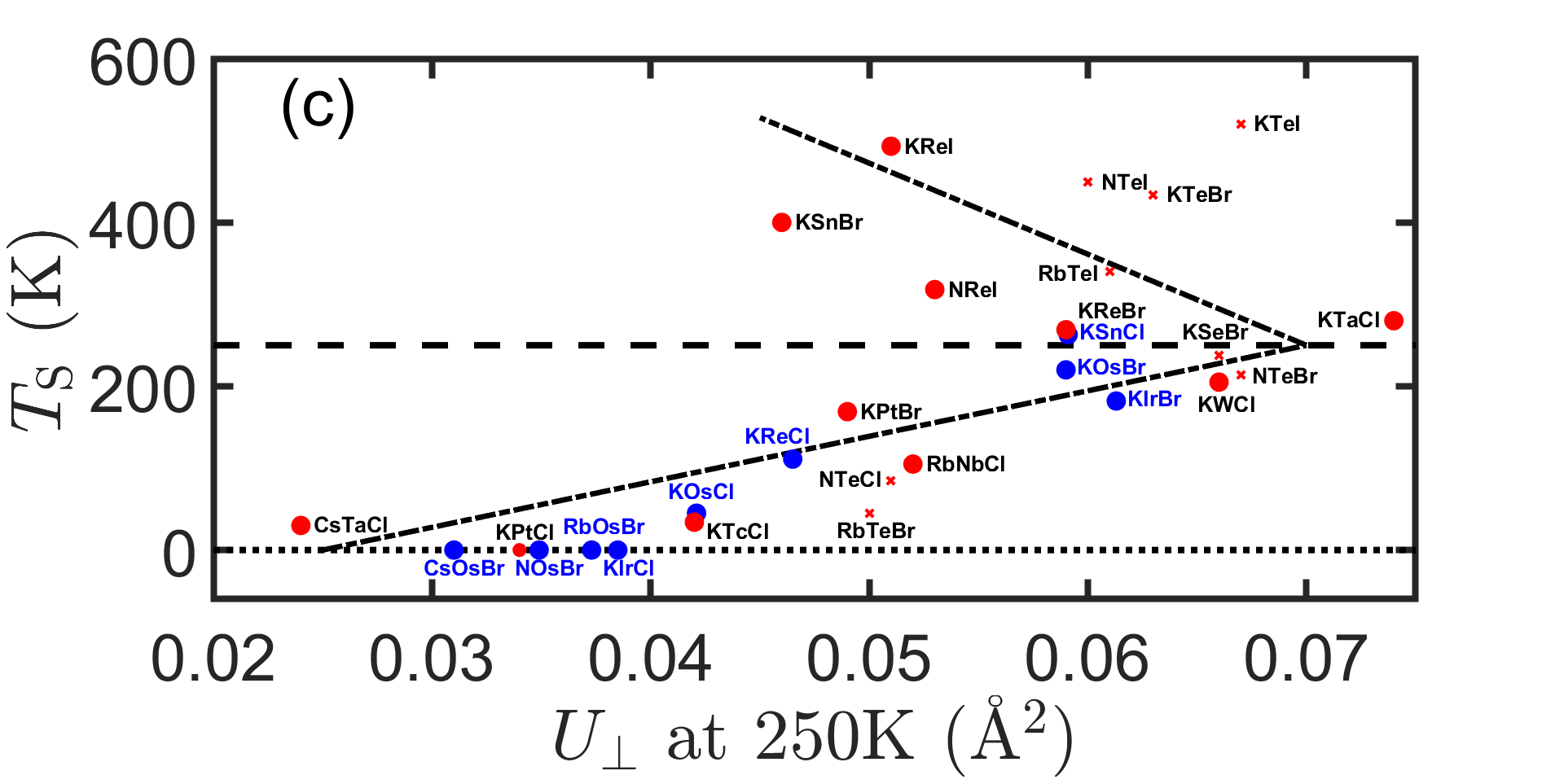} 
	\caption{\label{all_discussion} In all panels, the blue and red symbols refer to our own data and data from literature, respectively. Circle and cross symbols refer to $Me$ being a transition metal or Sn, or $Me=$Te, Pb, Se, respectively. The blue-shaded area in panel (a) highlights the former group. The chemical formula of $A_2MeX_6$ has been shortened to $AMeX$ with $N$=NH$_4$ for clarity. (a) Highest transition temperature $T_S$ from the cubic to a lower symmetry phase against Goldschmidt's tolerance factor $t$. The green thick line and the gray dashed-dotted line emphasize the linear behavior of $T_S$ against $t$ for $Me$ a $4d/5d$ transition metal and $Me=$Te, respectively. The black dashed line shows $T_S=0$.  (b)  Highest transition temperature $T_S$ against the square of the order parameter. The black dashed line is a guide to the eye. (c) Highest transition temperature $T_S$ against $U_{\rm perp}$ scaled to $T=250$K. The black dashed and dotted lines show $T_S=250$K and $T_S=0$K, respectively. The black dashed-dotted lines are a guide to the eye.}
\end{figure}

The highest transition temperature $T_S$ is plotted in Figure~\ref{all_discussion}(a) against the  tolerance factor $t$. Effective radii are taken from~\cite{Shannon76}, and for the ammonium group NH$_4$ from~\cite{Sidey76}. The blue-shaded area indicates antifluorite-type samples with $Me$ a $4d$/$5d$ transition metal ($Me=$Nb, Tc, Ru, Ta, W, Re, Os, Ir, Pt) or $Me=$Sn. When the tolerance factor is lower than one, the transition temperature increases linearly, as emphasized by the green line. It may appear surprising that the simple perovskite tolerance factor describes the occurrence of the rotational transitions so well. 
The empty $Me$ position in antifluorites compared to a perovskite does not invalidate such analysis; however, it causes an enhancement of the critical tolerance factor to $t_c \approx 1$ while perovskites can tolerate a larger mismatch up to $t_c \approx 0.95$. The empty $B'$ site thus enhances the instability, as the negative charge of the halide it not shared by two neighboring cations.
Following this analysis, it is obvious that Cs compounds of antifluorite-type do not exhibit the rotational transition. Note that the structural distortion in Cs$_2$TaCl$_6$ is of a different character~\cite{Ishikawa19}. 

Antifluorite-type compounds with $Me=$Te, Pb, and Se do not follow the seemingly linear behavior of $T_S$ against $t$ for $Me$ a transition metal. These clear deviations illustrate that the tabulated ionic radii for Te, Pb, and Se in a four-valent configuration are not applicable and cannot be used in the tolerance factor calculation, partially due to the character of the $Me-X$ bond being not essentially ionic. Furthermore, $A_2$Sb$X_6$ single crystals are reported to have a mixture of Sb$^{3+}$ and Sb$^{5+}$ rather than Sb$^{4+}$~\cite{Lawton66}. Note however that for compounds with $Me=$Te, $T_S$ follows also a linear tendency against $t$ suggesting an effective ionic radius of $\approx$0.67\AA~(instead of 0.97\AA) for Te$^{4+}$ in this material class. The Te-$X$ bond distances $d_{\rm calc,n}$ for several antifluorite compounds calculated with this corrected Te ionic radius agree better with the experimental values $d_{\rm exp}$, than those calculated with the literature value, see Tab.\ref{Te_bond_distance}.

\small
\begin{table}[h!]

	\centering
	\caption{\label{Te_bond_distance} The experimental bond distances Te-$X$ $d_{\rm exp}$ at room temperature are reported from literature for various samples and compared with the calculated bond distances $d_{\rm calc,n}$ and $d_{\rm calc,o}$. The latter are computed with the corrected and the original literature Te ionic radius, respectively. The standard ionic radius is used for the halide $X$.}
	\begin{indented}
		\lineup
		\item[]\begin{tabular}{@{}*{5}{l}}
			\br 
			Sample&Ref.& $d_{\rm exp}$(\AA)&$d_{\rm calc,n}$(\AA)&$d_{\rm calc,o}$(\AA) \cr	
			\mr		
		    (NH$_4$)$_2$TeCl$6$&\cite{Hazell66}&2.528(7)&\multirow{2}{*}{2.48}&\multirow{2}{*}{2.78} \cr
			Rb$_2$TeCl$_6$&\cite{Webster73}&2.525(5)&& \cr

			&&&& \cr
			K$_2$TeBr$_6$&\cite{Brown64}& 2.692(6)$^{\rm a}$&\multirow{5}{*}{2.63}&\multirow{5}{*}{2.93} \cr
			(NH$_4$)$_2$TeBr$6$&\cite{Das66}&2.695(5)&&\cr
			Rb$_2$TeBr$_6$&\cite{Abriel84,Abriel87}&2.689(3)&& \cr
			Cs$_2$TeBr$_6$&\cite{Das66}&2.695(5)&& \cr
										   &\cite{Abriel87}&2.703(2)&& \cr

			&&&& \cr
			K$_2$TeI$_6$&\cite{Syoyama72}&2.930$^{\rm a}$ &\multirow{3}{*}{2.87}&\multirow{3}{*}{3.17} \cr
(NH$_4$)$_2$TeI$6$&\cite{Abriel89}&2.934(3)$^{\rm a}$&&\cr
Rb$_2$TeI$_6$&\cite{Abriel82}&2.930(2)$^{\rm a}$&& \cr

			\br			
		\end{tabular}
		
\item[] $^{\rm a}$ Average bond distance for lower symmetry phase.	
	\end{indented}			
\end{table}
\normalsize

Because the structural distortions in the antifluorite-type compounds correspond to octahedra rotation and tilting, the absolute size of the order parameter can be defined as $\sqrt{\phi^2+\theta^2}$, where  $\phi$ and $\theta$  are the tilt angle around the $c$ axis and the rotation angle around the monoclinic $b_m$ axis, respectively. The transition temperature $T_S$, gauging the energy scale that triggers the structural transition, is plotted against the square of this order parameter in Figure~\ref{all_discussion}(b). Note that in a simple elastic model the square of the structural order parameter represents the elastic lattice energy (Hook law).  The proportionality between 
this order parameter square and the transitions temperature underlines the ubiquitous character of the 
rotation instability in this family of compounds. 
There are only two exceptions. According to the authors of~\cite{Syoyama72}, intensities of single-crystal XRD data of K$_2$TeI$_6$ could not be collected precisely. Therefore, the octahedra rotation angles calculated from the atomic positions may not be reliably determined, and the order parameter may be overestimated. With regards to Rb$_2$TeI$_6$~\cite{Abriel82}, the XRD data have been collected at room temperature, close to the cubic to tetragonal transition $T_S=340$~K. But additional structural transitions are expected at lower temperatures~\cite{Nakamura64} suggesting that the order parameter is strongly underestimated. The convincing scaling of $T_S$ with the tolerance factor and with $(\phi^2+\theta^2)$ illustrate the general character of the rotational instability in this broad class of materials.

As discussed previously, in the entire antifluorite-type family, $U_{\perp}$($X$) is much larger than $U_{\parallel}$($X$). The $U_{\perp}$ determined in the literature by means of X-ray or neutron diffraction at room temperature (unless stated otherwise), are plotted against the transition temperature $T_S$ in Figure~\ref{all_discussion}(c). All data are linearly scaled to T=250\,K. In the case of a monoclinic structure realized already at room temperature, $U_{\perp}$ corresponds to the largest eigenvalue of the ADP matrices. The closer $T_S$ is to 250\,K, indicated by the horizontal dotted line, the larger the transverse motion of the $X$ ADP is, which points to the role of phonon softening close to $T_S$. As expected, when  the structural transition $T_S$ is shifted far above or below room temperature, $U_{\perp}$, probing the structural instability at 250\,K, decreases. The dashed-dotted lines in Figure~\ref{all_discussion}(c) emphasize this behavior. This confirms our general conclusion that the large ADPs arise from the dynamics and not from local disorder.

\section{Conclusions}

Comprehensive single-crystal diffraction studies on K$_2$OsBr$_6$, K$_2$IrBr$_6$, K$_2$OsCl$_6$, K$_2$IrCl$_6$, Rb$_2$OsBr$_6$, and (NH$_4$)$_2$OsBr$_6$ reveal that none of these materials is suffering from significant chemical disorder or vacancies, except Cs$_2$OsBr$_6$,  which was found to exhibit a small amount of vacancies. The most remarkable structural feature visible in all the compounds concerns the general structural instability of the antifluorite-type family against rotation or tilting of the $Me$-halide octahedra, which causes two structural phase transitions in K$_2$OsBr$_6$ and K$_2$IrBr$_6$, and a single one in K$_2$OsCl$_6$. This instability is associated with soft phonon modes that cover a large part of the Brillouin zone due to a reduced dispersion. In consequence the impact of these soft phonons is remarkably high resulting in strongly anisotropic ADPs at the halide site. However the large transversal ADP $U_{\perp}$ of the halide that senses octahedron rotations strongly shrinks upon cooling underlining its dynamic character. Refinements of anharmonic ADPs yield some improvement in the description of the Bragg intensities, but these improvements are small. The anharmonic treatment of the four compounds K$_2$OsBr$_6$, K$_2$IrBr$_6$, K$_2$OsCl$_6$, and K$_2$IrCl$_6$ precludes local disorder in the cubic phase as origin of the non cubic crystal-field seen in the iridates, but absent in the osmate materials. In these antifluorite-type materials, the characters of the instability and of the structural phase transitions remain essentially displacive and the large values of the ADPs arise from the dynamics. 

Comparing our structural data with studies on numerous other $A_2MeX_6$ compounds, we find a convincing scaling of the upper transition temperature $T_S$ with the simple tolerance factor known for perovskites. The empty $B'$ sites just imply a shift of the critical tolerance factor to $t_c \approx 1$. Also the scaling between the transition temperature and the square sum of the low-temperature octahedron rotation angles underlines the general character of the rotational instability in this broad class of materials.

\ack
	We acknowledge support by the DFG (German Research Foundation) via Project No. 277146847-CRC 1238 (Subproject A02).

\appendix

\section{Correlation plots}
\label{correlation_plot}

\begin{figure}[h!]

			\includegraphics[width=0.49\columnwidth]{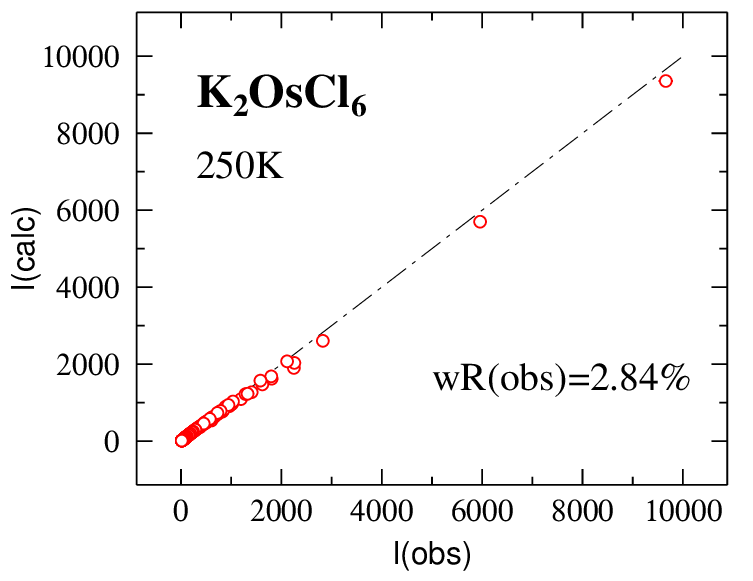}
			\includegraphics[width=0.49\columnwidth]{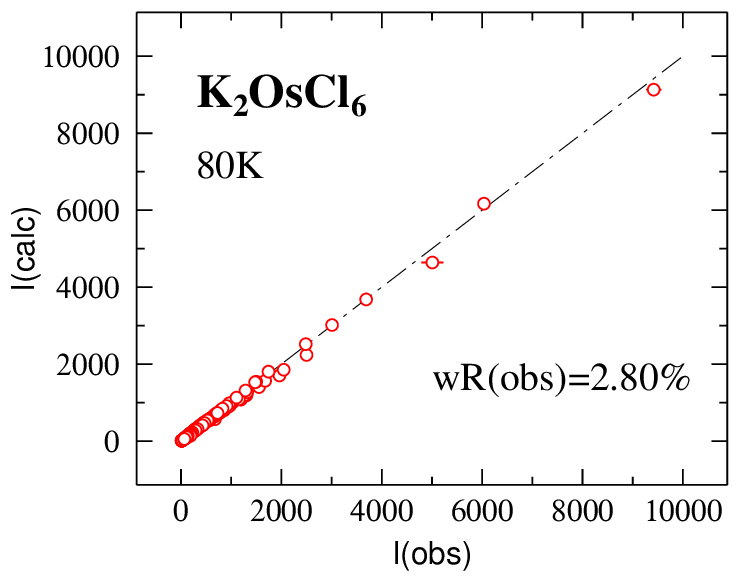}				
	\caption{\label{IoIc} Calculated (I(calc)) versus measured (I(obs)) integrated intensities plots of all unique reflections listed in Table~\ref{Data_set_all}, for K$_2$OsCl$_6$. All model refinements have been carried out with space group $Fm\bar{3}m$ and anharmonic ADPs. The black dashed-dotted line shows the ideal relation I(calc)=I(obs).}		
\end{figure}

\begin{figure}[h!]				
			\includegraphics[width=0.49\columnwidth]{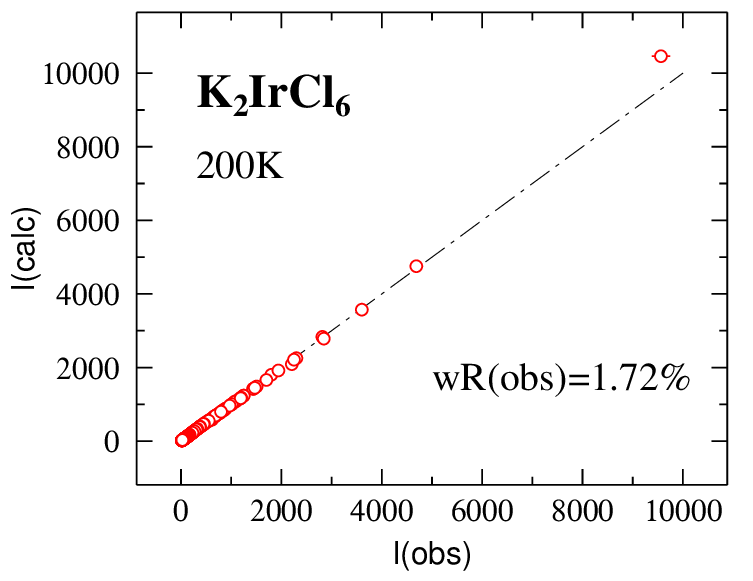}
			\includegraphics[width=0.49\columnwidth]{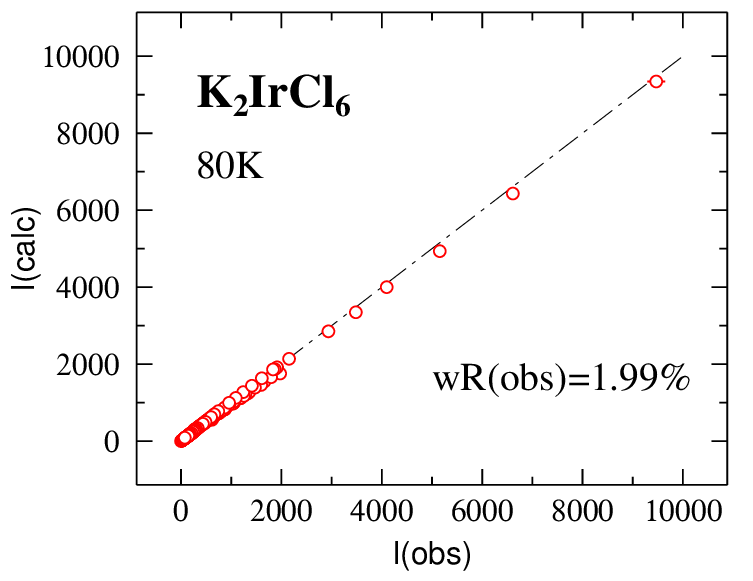} 		
			\includegraphics[width=0.49\columnwidth]{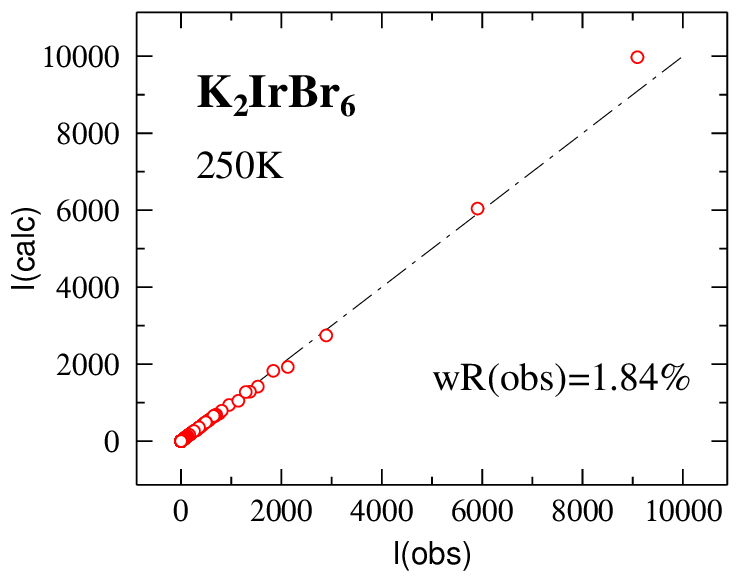} 				
			\includegraphics[width=0.49\columnwidth]{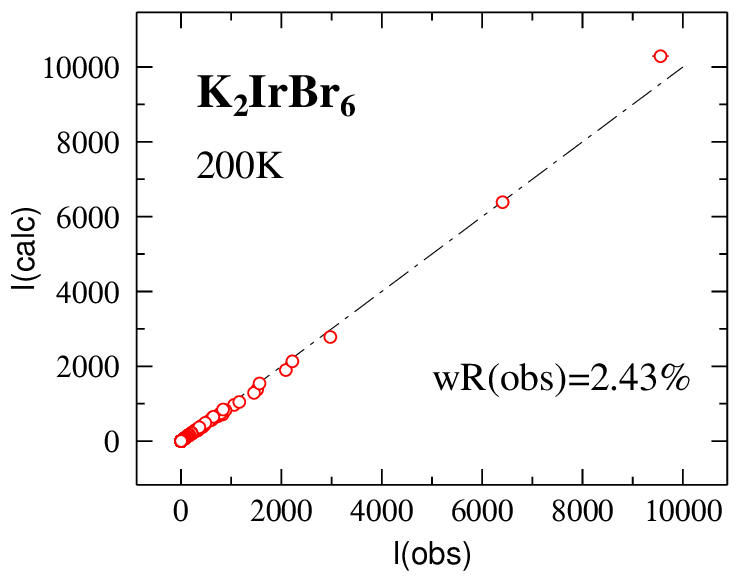} 			
			\includegraphics[width=0.49\columnwidth]{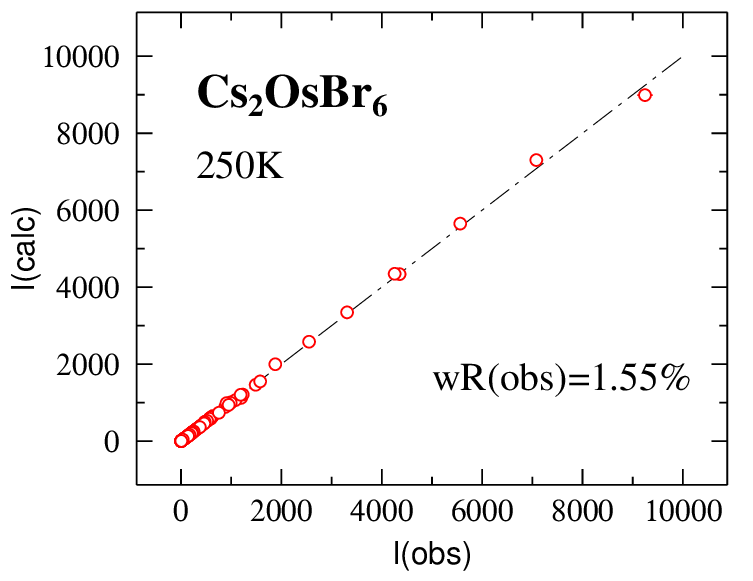} 					
			\includegraphics[width=0.49\columnwidth]{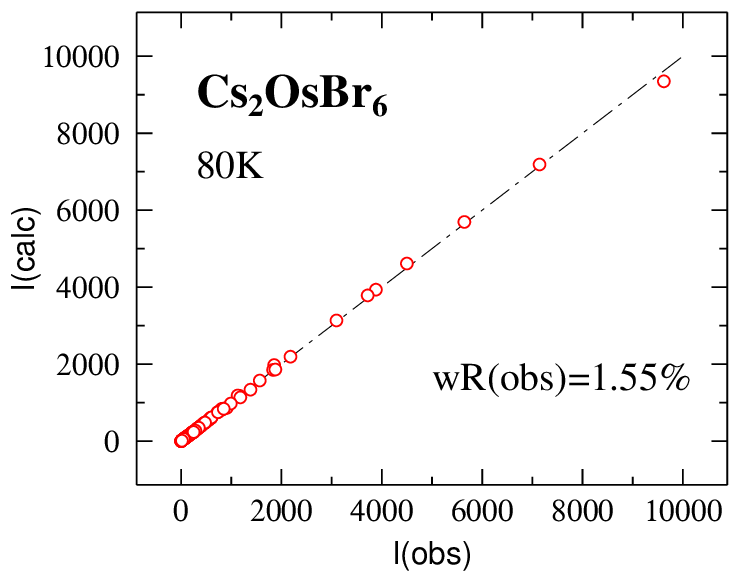} 
		    \includegraphics[width=0.49\columnwidth]{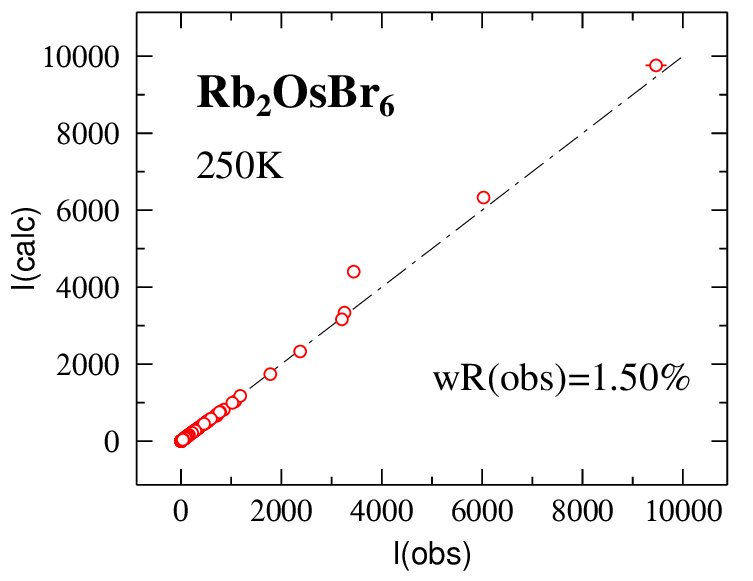} 
			\includegraphics[width=0.49\columnwidth]{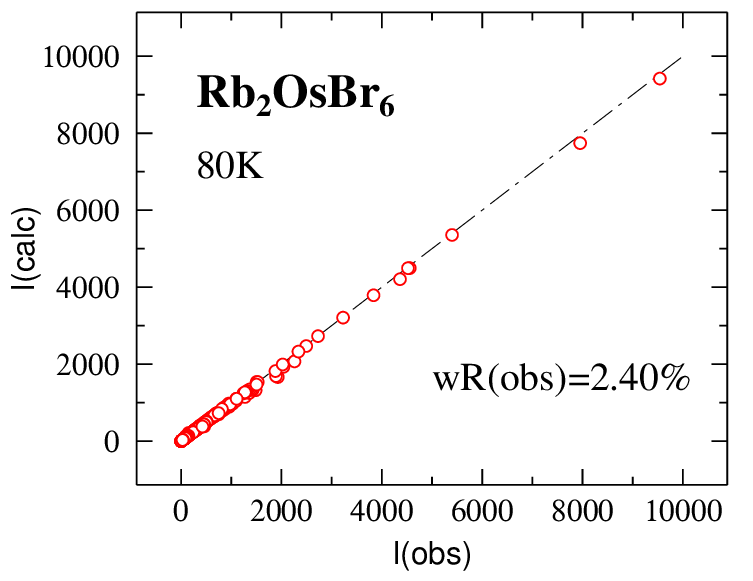}  	
		   \includegraphics[width=0.49\columnwidth]{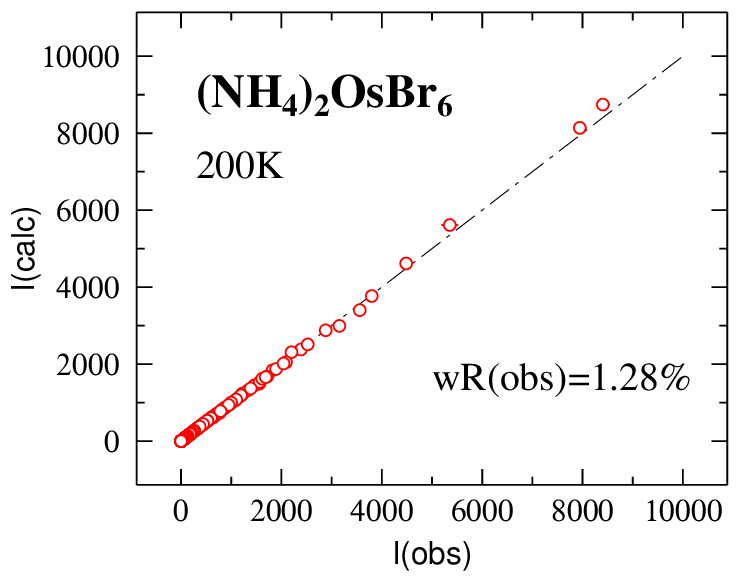} 
	    	\includegraphics[width=0.49\columnwidth]{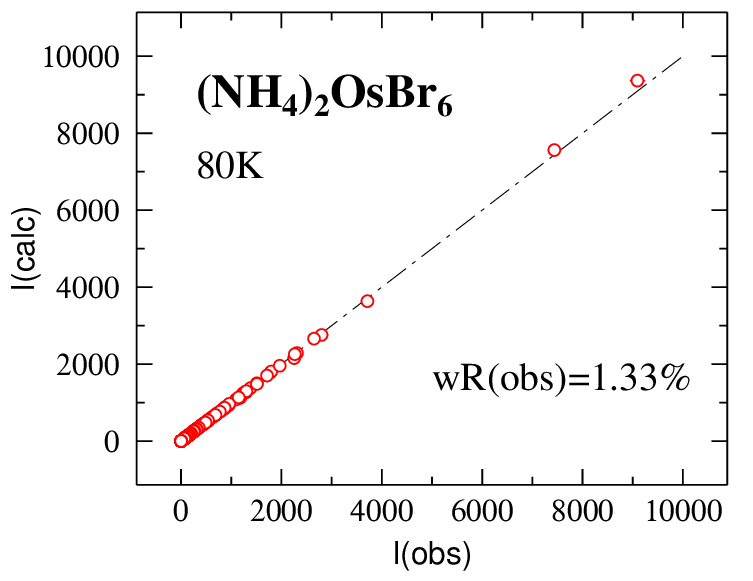} 	
	\caption{\label{IoIc_2} Calculated (I(calc)) versus measured (I(obs)) integrated intensities plots of all unique reflections listed in Table~\ref{Data_set_all}, for samples K$_2$IrCl$_6$, K$_2$IrBr$_6$, Cs$_2$OsBr$_6$, Rb$_2$OsBr$_6$, and (NH$_4$)$_2$OsBr$_6$. All model refinements have been carried out with space group $Fm\bar{3}m$ and anharmonic ADPs. The black dashed-dotted line shows the ideal relation I(calc)=I(obs).}		
\end{figure}

\begin{figure}[h!]
	\raggedright
				\includegraphics[width=0.49\columnwidth]{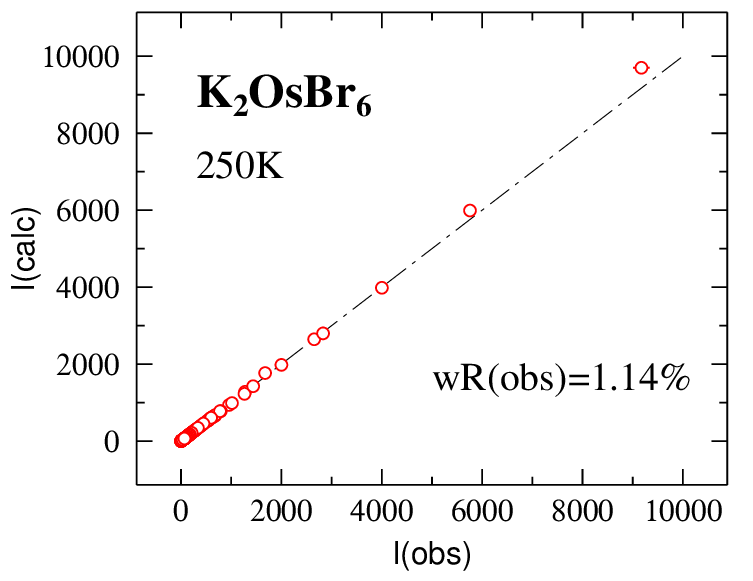} 
				\includegraphics[width=0.49\columnwidth]{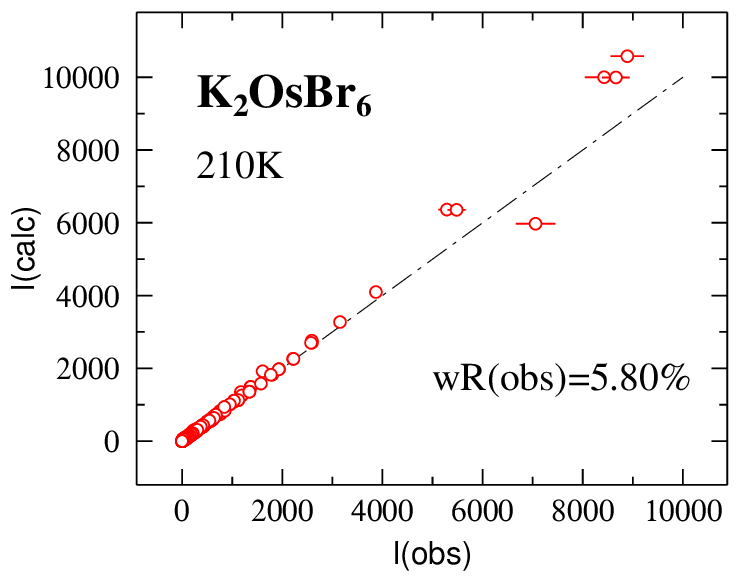} 
		 		\includegraphics[width=0.49\columnwidth]{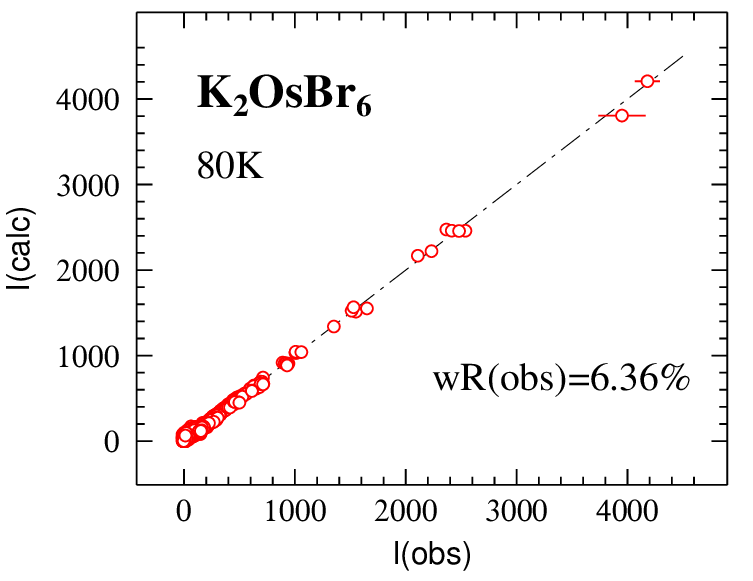} 								
	\caption{\label{IoIc_3} Calculated (I(calc)) versus measured (I(obs)) integrated intensities plots of all unique reflections listed in Table~\ref{Data_set_all}, for K$_2$OsBr$_6$. Model refinements have been performed at 250K with space group $Fm\bar{3}m$ and anharmonic ADPs, and at 210~K and 80~K with space groups $P4/mnc$ and $P2_1/n$, and harmonic ADPs, respectively. The black dashed-dotted line shows the ideal relation I(calc)=I(obs).}		
\end{figure}

In order to further asses the quality of all the refinements discussed in the main text, we report in Figure~\ref{IoIc},  Figure~\ref{IoIc_2} and Figure~\ref{IoIc_3} the calculated I(calc) versus measured I(obs) integrated intensities plots of all unique reflections listed in Table~\ref{Data_set_all}. All plots correspond to refinements performed with space group $Fm\bar{3}m$ and anharmonic ADPs, with the exception of K$_2$OsBr$_6$ at 210K and 80K, where refinements have been carried out with space groups $P4/mnc$ and $P2_1/n$, respectively, and with harmonic ADPs.


\clearpage
\newpage
\section*{References}


\providecommand{\newblock}{}

\end{document}